\documentclass[aps,prl,showpacs]
{revtex4}

\usepackage[utf8]{inputenc} 
\usepackage{amsmath}
\usepackage{amssymb}
\usepackage{mathrsfs}
\usepackage{bm}
\usepackage{color}

\usepackage{graphicx}
\DeclareGraphicsExtensions{.pdf,.jpg,}

\usepackage{caption}

\newcommand{\ket}[1]{| #1 \rangle}
\newcommand{\bra}[1]{\langle #1 |}

\newcommand{\tr}{\text{Tr}}
\newcommand{\q}[1]{\vec{#1}\cdot\vec{\sigma}}

\begin{document}

\author{Bruno Mera$^{1,2,4}$, Chrysoula Vlachou$^{3,4}$, Nikola Paunkovi\'c$^{3,4}$ 
and V\'{\i}tor R. Vieira$^{1,2}$}
\affiliation{$^1$ CeFEMA, Instituto Superior
T\'ecnico, Universidade de Lisboa, Av. Rovisco Pais, 1049-001 Lisboa, Portugal}
\affiliation{$^2$  Departamento de F\'{\i}sica, Instituto Superior
T\'ecnico, Universidade de Lisboa, Av. Rovisco Pais, 1049-001 Lisboa, Portugal}
 \affiliation{$^3$ Departamento de Matemática, Instituto Superior Técnico, Universidade de Lisboa, Av. Rovisco Pais, 1049-001 Lisboa, Portugal}
\affiliation{ $^4$ Instituto de Telecomunica\c{c}\~oes, 1049-001 Lisbon, Portugal}
\title{
The Uhlmann connection in 
fermionic systems undergoing phase transitions
}
\begin{abstract}
We study the behavior of the Uhlmann connection in systems of fermions undergoing phase transitions. In particular, we analyze some of the paradigmatic cases of topological insulators and superconductors in 1D, as well as the BCS theory of superconductivity in 3D. We show that the Uhlmann connection signals phase transitions in which the eigenbasis of the state of the system changes. Moreover, using the established fidelity approach and the study of the edge states, we show the absence of thermally driven phase transitions in the case of topological insulators and superconductors. We clarify what is the relevant parameter space associated with the Uhlmann connection so that it signals the existence of order in mixed states. In addition, the study of Majorana modes at finite temperature opens the possibility of applications in realistic stable quantum memories. Finally, the analysis of the different behavior of the BCS model and the Kitaev chain, with respect to the Uhlmann connection, suggested that in realistic scenarios the gap of topological superconductors could also, generically, be temperature-dependent.   

\end{abstract}
\maketitle

\section{Introduction}
\label{sec:introduction}
Topological phases of matter are a subject of active research during the last decades, as they constitute a whole new paradigm in condensed matter physics. In contrast to the well-studied standard quantum phases of matter, described by local order parameters (see for example Anderson's classification~\cite{and:07}), the ground states of topological systems are globally characterized by topological invariants~\cite{tknn:82,ber:84,zen:che:zho:wen:15,ando:13}. Hamiltonians of gapped systems with different topological orders cannot be smoothly transformed from one into the other unless passing through a gap-vanishing region of criticality. In particular, insulators and superconductors with an energy gap exhibit topological orders and are classified according to the symmetries that their Hamiltonians possess~\cite{kit:09, sch:ryu:fur:lud:08}, namely Time Reversal, Particle-Hole and Chiral symmetry. As opposite to the standard Landau symmetry breaking theory of quantum phase transitions (PTs), in topological PTs the symmetries of the Hamiltonian are not violated. For a topological PT to occur, that is a gapped state of the system to be deformed in another gapped state in a different topological class, the energy gap has to close. In other words, the quantum state of the system undergoing a topological PT is gapless. A manifestation of the topological order of a system is the presence of robust symmetry-protected edge states on the boundary between two distinct topological phases, as predicted by the bulk-to-boundary principle~\cite{ryu:hat:02}. 

A question that naturally arises is whether there is any kind of topological order at finite temperatures, and different approaches have been used to tackle this problem~\cite{riv:viy:del:13,viy:riv:del:12}. One of the most promising approaches is based on the work of Uhlmann~\cite{uhl:89}, who extended the notion of geometrical phases from pure states to density matrices. The concept of the Uhlmann holonomy, and the quantities that can be derived from it, were used to infer PTs at finite temperatures~\cite{hub:93,viy:riv:del:14, viy:riv:del:2d:14,zho:aro:14,pau:vie:08,viy:riv:del:15}. Nevertheless, the physical meaning of these quantities and their relevance to the observable properties of the corresponding systems stay as an interesting open question~\cite{sjo:15, kem:que:mor:16, bud:die:15}. There exist several proposals for the observation of the Uhlmann geometric phase~\cite{tid:sjo:03, abe:kul:sjo:oi:07, viy:riv:gas:wal:fil:del:16} and an experimental realization has been reported in~\cite{uhl:pha:exp:11}. 

Information-theoretic quantities such as entanglement measures~\cite{ham:ion:zan:05, ham:zha:haa:lid:08, hal:ham:12} and the fidelity, a measure of distinguishability between two quantum states~\cite{aba:ham:zan:08, zan:pau:06, zha:zho:09, oli:sac:14, pau:sac:nog:vie:dug}, were extensively used in the study of PTs. Whenever there is a PT, the density matrix of a system changes significantly and therefore, a sudden drop of the fidelity $F(\rho,\sigma)\equiv \tr\sqrt{\sqrt{\rho}\sigma\sqrt{\rho}}$, signals out this change. The fidelity, is closely related to the Uhlmann connection, through the Bures metric~\cite{zan:ven:gio:07}. Therefore, they can both be used to infer the possibility of PTs, as in~\cite{pau:vie:08} (for the pure-state case of the Berry phase, see~\cite{car:pac:05,reu:har:ple:07}).\\
		
We analyze the behavior of the fidelity and the Uhlmann connection associated to thermal states in fermionic systems. We consider the space consisting of the parameters of the Hamiltonian and the temperature, as it provides a physically sensible base space for the principal bundle, describing the amplitudes of the density operator. We study paradigmatic models of 1D topological insulators (TIs) (Creutz Ladder~\cite{cre:99,ber:pat:ami:del:09} and Su-Schrieffer-Heeger~\cite{su:sch:hee:79} models) and superconductors (TSCs) (Kitaev chain~\cite{kit:cha:01}) with chiral symmetry. We conclude that the effective temperature only smears out the topological features exhibited at zero temperature, without causing a thermal PT. We also analyze the BCS model of superconductivity~\cite{bar:coo:sch:57}, previously studied in~\cite{pau:vie:08}, by further identifying the significance of thermal and purely quantum contributions to PTs, using the fidelity and the Uhlmann connection. In contrast to the aforementioned nontrivial topological systems, both quantities indicate the existence of thermal PTs.\\

This Letter is organized as follows: first, we elaborate on the relationship between the fidelity and the Uhlmann connection and motivate their use in inferring PTs, both at zero and finite temperatures. In the following section, we present our results on the fidelity and the Uhlmann connection for the aforementioned systems and discuss the possibility of temperature driven PTs. In the last section we summarize our conclusions and point out possible directions of future work.

\section{Fidelity and the Uhlmann Parallel Transport}
Given a Hilbert space, one can consider the set of density matrices with full rank (e.g. thermal states) and the associated set of amplitudes (generalization to the case of the sets of singular density matrices with fixed rank is straightforward). For a state $\rho$, an associated amplitude $w$ satisfies $\rho = ww^\dagger$. Thus, there exists a unitary (gauge) freedom in the choice of the amplitude, since both $w$ and $w' = wU$, with $U$ being an arbitrary unitary, are associated to the same $\rho$. Two amplitudes $w_1$ and $w_2$, corresponding to states $\rho_1$ and $\rho_2$, respectively, are said to be \emph{parallel} in the Uhlmann sense if and only if they minimize the Hilbert- Schmidt distance $||w_1 - w_2|| = \sqrt{\tr [(w_1 - w_2)^\dagger (w_1 - w_2)]}$, induced by the inner product $\langle w_1,w_2\rangle =  \tr(w_1^\dagger w_2)$. The condition of parallelism turns out to be equivalent to maximizing $\text{Re}\langle w_1,w_2\rangle$, since $||w_1 - w_2||^2 = 2(1 - \text{Re}\langle w_1,w_2\rangle)$. By writing $w_i=\sqrt{\rho}_i U_i$, $i=1,2$, where the $U_i$'s are unitary matrices, we get
\begin{align}
\text{Re}\langle w_1,w_2\rangle & \leq  |\langle w_1,w_2\rangle| =|\tr \left(w_2^\dagger w_1\right)|\nonumber\\
&= |\tr\left( U_2^\dagger \sqrt{\rho_2}\sqrt{\rho_1}U_1\right)| \nonumber \\
		& = |\tr\left(|\sqrt{\rho_2}\sqrt{\rho_1}|U U_1 U_2^\dagger\right)| \nonumber \\ & \leq \tr\left(|\sqrt{\rho_2}\sqrt{\rho_1}|\right)\nonumber\\
	& = \tr\sqrt{\sqrt\rho_1 \rho_2 \sqrt\rho_1} = F(\rho_1,\rho_2),
\end{align}
where $U$ is the unitary associated to the polar decomposition of $\sqrt{\rho_2}\sqrt{\rho_1}$, and the penultimate step is the Cauchy-Schwartz inequality. Hence, the equality holds if and only if $U (U_1U_2^\dagger) =I$. Note that in this case, also the first equality holds, and we have $\text{Re}\langle w_1,w_2\rangle\ = \langle w_1,w_2\rangle \in \mathbb R^+$, which provides yet another interpretation of the Uhlmann parallel transport condition as a generalization of the Berry pure-state connection: the phase, given by $\Phi_U = \arg \langle w_1,w_2\rangle$, is trivial, i.e., zero.

Given a curve of density matrices $\gamma:[0,1] \ni t\mapsto \rho(t)$ and the initial amplitude $w(0)$ of $\rho (0)$, the Uhlmann parallel transport gives a unique curve of amplitudes $w(t)$ with the property that $w(t)$ is parallel to $w(t+\delta t)$ for an infinitesimal $\delta t$ (the horizontal lift of $\gamma$). The length of this curve of amplitudes, according to the metric induced by the Frobenius inner product, is equal to the length, according to the Bures metric (which is the infinitesimal version of the Bures distance $D_{B}(\rho_1,\rho_2)^2=2[1-F(\rho_1,\rho_2)]$), of the corresponding curve $\gamma$ of the density matrices. This shows the relation between the Uhlmann connection and the fidelity (for details, see for example~\cite{uhl:11, uhl:89}).

We see that the ``Uhlmann factor'' $U$, given by the polar decomposition $\sqrt{\rho (t+\delta t)}\sqrt{\rho (t)} = |\sqrt{\rho (t+\delta t)}\sqrt{\rho (t)}| U$, characterizes the Uhlmann parallel transport. For two close points $t$ and $t + \delta t$, if the two states $\rho (t)$ and $\rho (t+\delta t)$ belong to the same phase, one expects them to almost commute, resulting in the Uhlmann factor being approximately equal to the identity, $\sqrt{\rho (t+\delta t)}\sqrt{\rho (t)} \approx |\sqrt{\rho (t+\delta t)}\sqrt{\rho (t)}|$. On the other hand, if the two states belong to two different phases, one expects them to be drastically different (confirmed by the fidelity approach), both in terms of their eigenvalues and/or eigenvectors, potentially leading to nontrivial $U \neq I$ (see the previous study on the Uhlmann factor and the finite-temperature PTs for the case of the BCS superconductivity~\cite{pau:vie:08}). To quantify the difference between the Uhlmann factor and the identity, and thus the nontriviality of the Uhlmann connection, we consider the following quantity:
\begin{align}
\Delta (\rho(t),\rho(t+\delta t)):=& F(\rho(t),\rho(t+\delta t)) \nonumber \\
&-\tr(\sqrt{\rho(t+\delta t)}\sqrt{\rho(t)}). 
\end{align}
Note that $\Delta = \tr \big[|\sqrt{\rho(t+\delta t)}\sqrt{\rho(t)}|(I-U)\big]$. When the two states are from the same phase we have $\rho (t) \approx \rho (t+ \delta t)$, and thus $\Delta  \approx 0$. Otherwise, if the two states belong to different phases, and the Uhlmann factor is nontrivial, we have $\Delta \neq  0$. Thus, sudden departure of $\Delta$ from zero (for $\delta t << 1$) signals the points of PTs. Since both the Uhlmann parallel transport and the fidelity give rise to the same metric (the Bures metric), the non-analyticity of $\Delta$ is accompanied by the same behaviour of the fidelity. Note that the other way around is not necessarily true: in case the states commute with each other and differ only in their eigenvalues, the Uhlmann connection is trivial, and thus $\Delta=0$.

In order for the Uhlmann connection and the fidelity to be in tune, they must be taken over the same base space. In previous studies~\cite{viy:riv:del:14}, an Uhlmann connection in 1D translationally invariant systems was considered. The base space considered is the momentum space and the density matrices are of the form $\{\rho_k:=e^{-\beta H(k)}/Z: k\in \mathcal B \}$, where $H(k)=E(k)\vec{n}(k)\cdot \vec{\sigma}/2$ and $\mathcal B$ is the first Brillouin zone. Since we are in 1D, there exists no curvature and hence the holonomy along the momentum space cycle becomes a topological invariant (depends only on the homotopy class of the path). It was found that the Uhlmann geometric phase $\Phi_U(\gamma_c)$ along the closed curve given by $\gamma_c(k)=\rho_k$, changes abruptly from $\pi$ to $0$ after some ``critical'' temperature $T_U$. Namely, the Uhlmann phase is given by 
\begin{align}
\Phi_U(\gamma_c)&=\arg\tr\{w(-\pi)^{\dagger}w(\pi)\} \nonumber\\
&=\arg\tr\{\rho_{\pi}U(\gamma_c)\},
\end{align}
where $w(k)$ is the horizontal lift of the loop of density matrices $\rho_{k}$, and $U(\gamma_c)$ is the Uhlmann holonomy along the first Brillouin zone. This temperature, though, is not necessarily related to a physical quantity that characterizes a system's phase. It might be the case that the Uhlmann phase is trivial, $\Phi_U(\gamma_c) = 0$, while the corresponding holonomy is not, $U(\gamma_c)\neq I$. For the systems studied in~\cite{viy:riv:del:14}, the Uhlmann holonomy is a smooth function of the temperature and is given, in the basis in which the chiral symmetry operator is diagonal, by:
\begin{align}
U(\gamma_c) = \exp \Big\{-\frac{i}{2} \int_{-\pi}^{\pi} \left[1 - \text{sech} \left(\frac{E(k)}{2T}\right)\right]\frac{\partial \varphi}{\partial k}dk\  \sigma_z\Big\},
\label{eq:hol}	
\end{align}
where $\varphi (k)$ is the polar angle coordinate of the vector $\vec{n}(k)$ lying on the equator of the Bloch sphere. Note that $\lim_{T\rightarrow 0} U(\gamma_c) = e^{-i\nu\pi\sigma_z}$, with the Berry phase being $\Phi_B = \lim_{T\rightarrow 0} \Phi_U = \nu\pi$, and $\nu$ the winding number. While in this case the Uhlmann phase suffers from an abrupt change (step-like behaviour), the Uhlmann holonomy is smooth and there is no PT-like behaviour. 

In the paradigmatic case of the quantum Hall effect, at $T=0$, the Hall conductivity is quantized in multiples of the first Chern number of a vector bundle in momentum space through several methods. For example, one can use linear response theory or integrate the fermions to obtain the effective action of an external $\text{U}(1)$ gauge field. The topology of the bands appears, thus, in the response of the system to an external field. It is unclear, though, that the former mathematical object, the Uhlmann geometric phase along the cycle of the 1D momentum space, has an interpretation in terms of the response of the system. In order to measure this Uhlmann geometric phase, one would have to be able to change the quasi-momentum of a state in an adiabatic way. In realistic setups, the states at finite temperatures are statistical mixtures over all momenta, such as the thermal states considered, and realizing closed curves of states $\rho_k$ with precise momenta changing in an adiabatic way seems a tricky task. The fidelity computed in our Letter though, refers to the change of the system's {\em overall} state, with respect to its parameters (controlled in the laboratory much like an external gauge field), and is related to an, \textit{a priori}, physically relevant geometric quantity, the Uhlmann factor $U$. The quantity $\Delta=\tr \big[|\sqrt{\rho(t+\delta t)}\sqrt{\rho(t)}|(I-U)\big]$ contains information concerning the Uhlmann factor, since $U=U(t+\delta t) U^\dagger(t)$, where $U(t)=T\exp\left\{-\int_{0}^{t} \mathcal{A}(d\rho/ds)ds\right\}$ is the parallel transport operator and $\mathcal{A}$ is the Uhlmann connection differential $1$-form (for details, see~\cite{chr:jam:12}).

\section{The results}
In our analysis we probe the fidelity and $\Delta$ with respect to the parameters of the Hamiltonian describing the system and the temperature, independently. We perform this analysis for paradigmatic models of TIs (SSH and Creutz ladder) and TSCs (Kitaev Chain) in 1D. We analytically calculate the expressions for the fidelity and $\Delta$, for thermal states $\rho=e^{-\beta H}/Z$, where $\beta$ is the inverse temperature (see SM1 for the details of the derivation). We use natural units: $\hbar=k_{\text{B}}=1$.\\
\indent Here we focus on the Creutz Ladder model, while the results for SSH and the Kitaev chain are presented in SM2, since they are qualitatively the same. The Hamiltonian for the Creutz Ladder model~\cite{cre:99,ber:pat:ami:del:09} is given by
\begin{align}
\mathcal{H}=& -\sum_{i\in\mathbb{Z}}K\left(e^{-i\phi}a_{i+1}^{\dagger}a_{i}+e^{i\phi}b_{i+1}^{\dagger}b_{i}\right)\nonumber\\
&+K(b_{i+1}^{\dagger}a_{i}+a_{i+1}^{\dagger}b_{i})+Ma_{i}^\dagger b_i +\text{H.c.},	
\end{align}
where $a_i,b_i$, with $i\in\mathbb{Z}$, are fermion annihilation operators, $K$ and $M$ are hopping amplitudes (horizontal/diagonal and vertical, respectively) and $e^{i\phi}$ is a phase factor associated to a discrete gauge field. We take $2K=1$, $\phi=\pi/2$. Under these conditions, the system is topologically nontrivial when $M<1$ and trivial when $M>1$. 
Given two close points $(M,T)$ and $(M',T')=(M+\delta M, T+\delta T)$, we compute $F(\rho,\rho')$ and $\Delta(\rho,\rho')$ between the states $\rho=\rho(M,T)$ and $\rho'=\rho(M',T')$. To distinguish the contributions due to the change of Hamiltonian's parameter and the temperature, we consider the cases $\delta T = 0$ and $\delta M = 0$, respectively, see Fig.~\ref{fig:fid}.

\begin{widetext}

\begin{figure}[h!]
\begin{minipage}{0.33\textwidth}
\includegraphics[width=0.8\textwidth,height=0.6\textwidth]{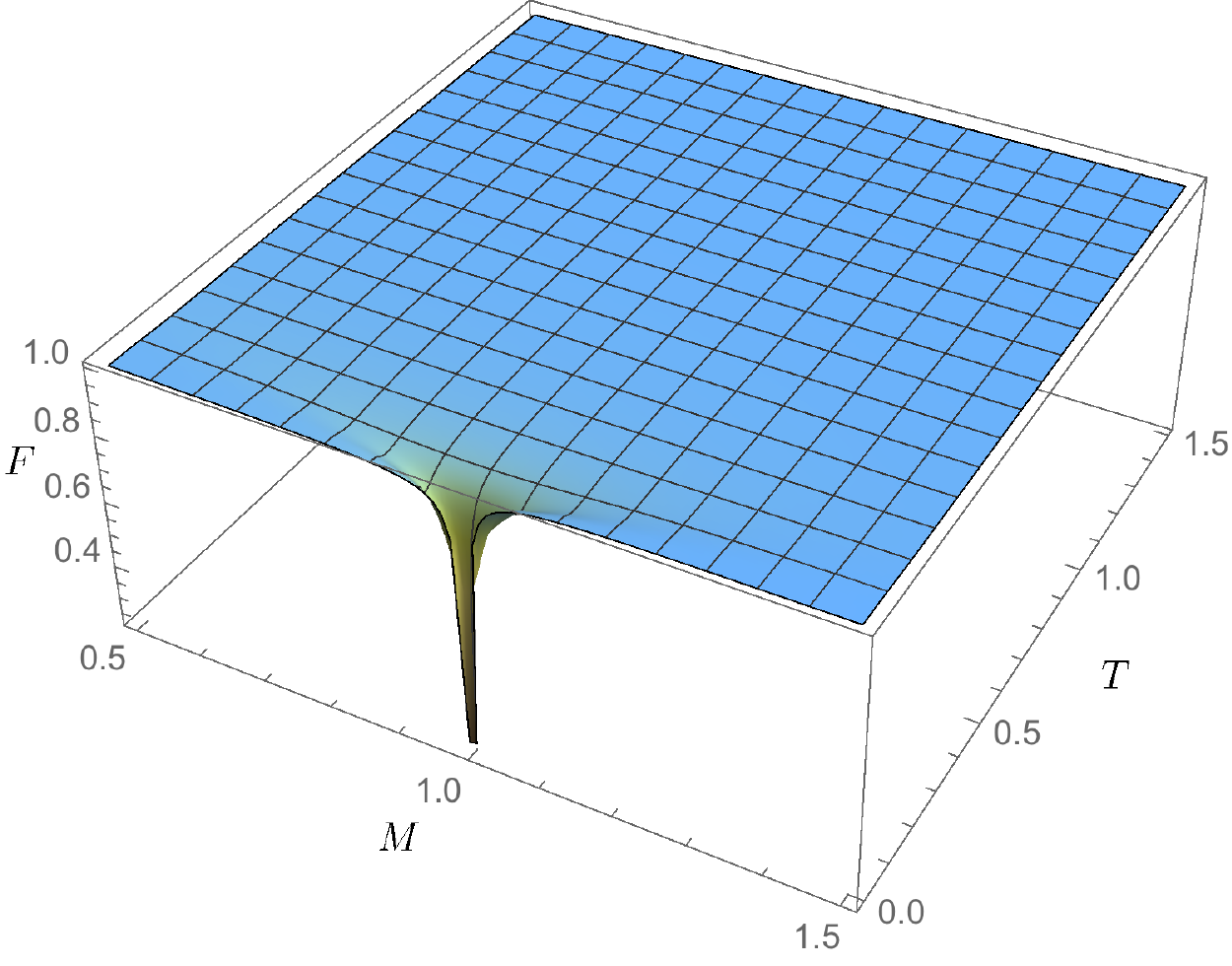}
\end{minipage}%
\begin{minipage}{0.33\textwidth}
\includegraphics[width=0.8\textwidth,height=0.6\textwidth]{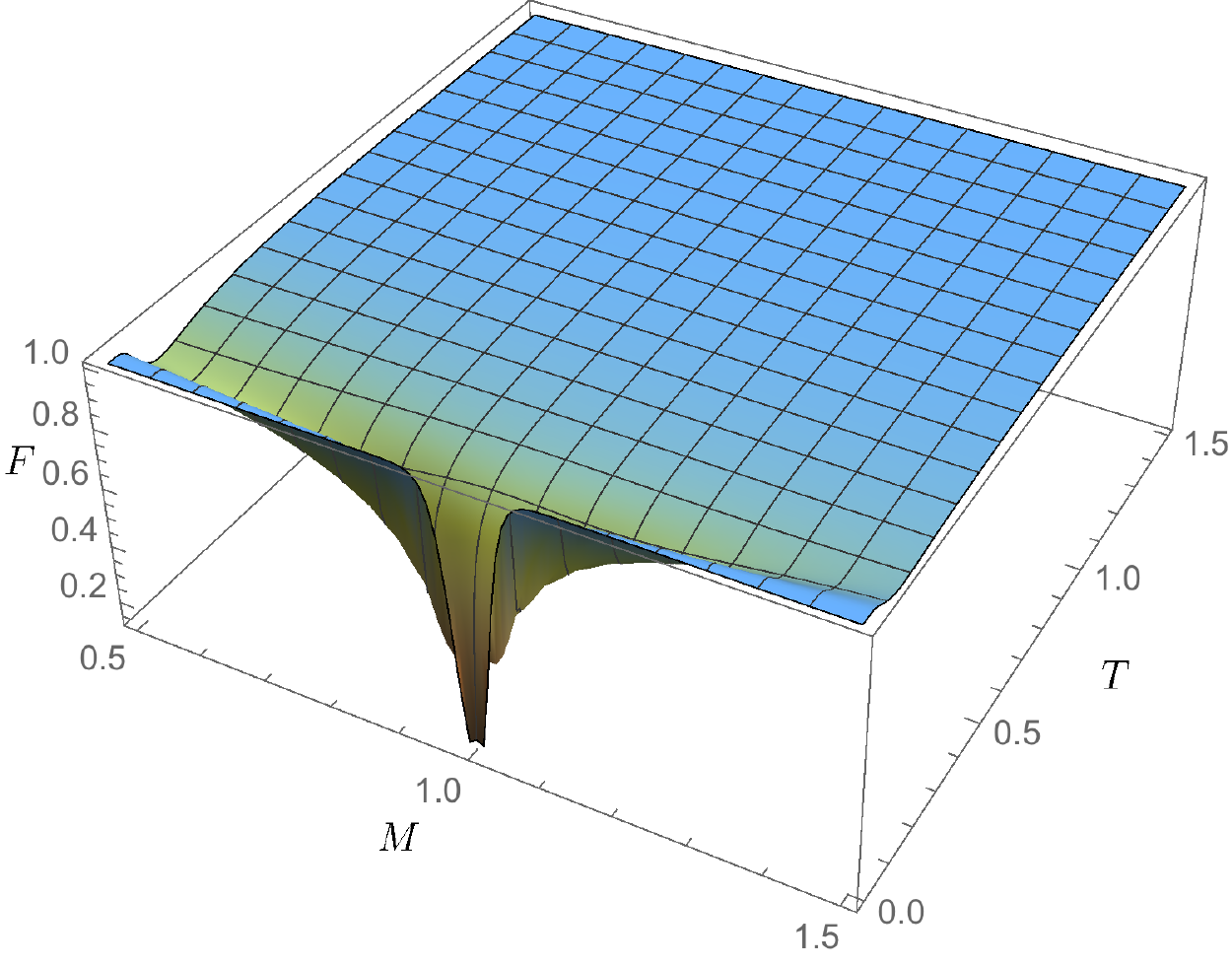}
\end{minipage}
\begin{minipage}{0.33\textwidth}
\includegraphics[width=0.8\textwidth,height=0.6\textwidth]{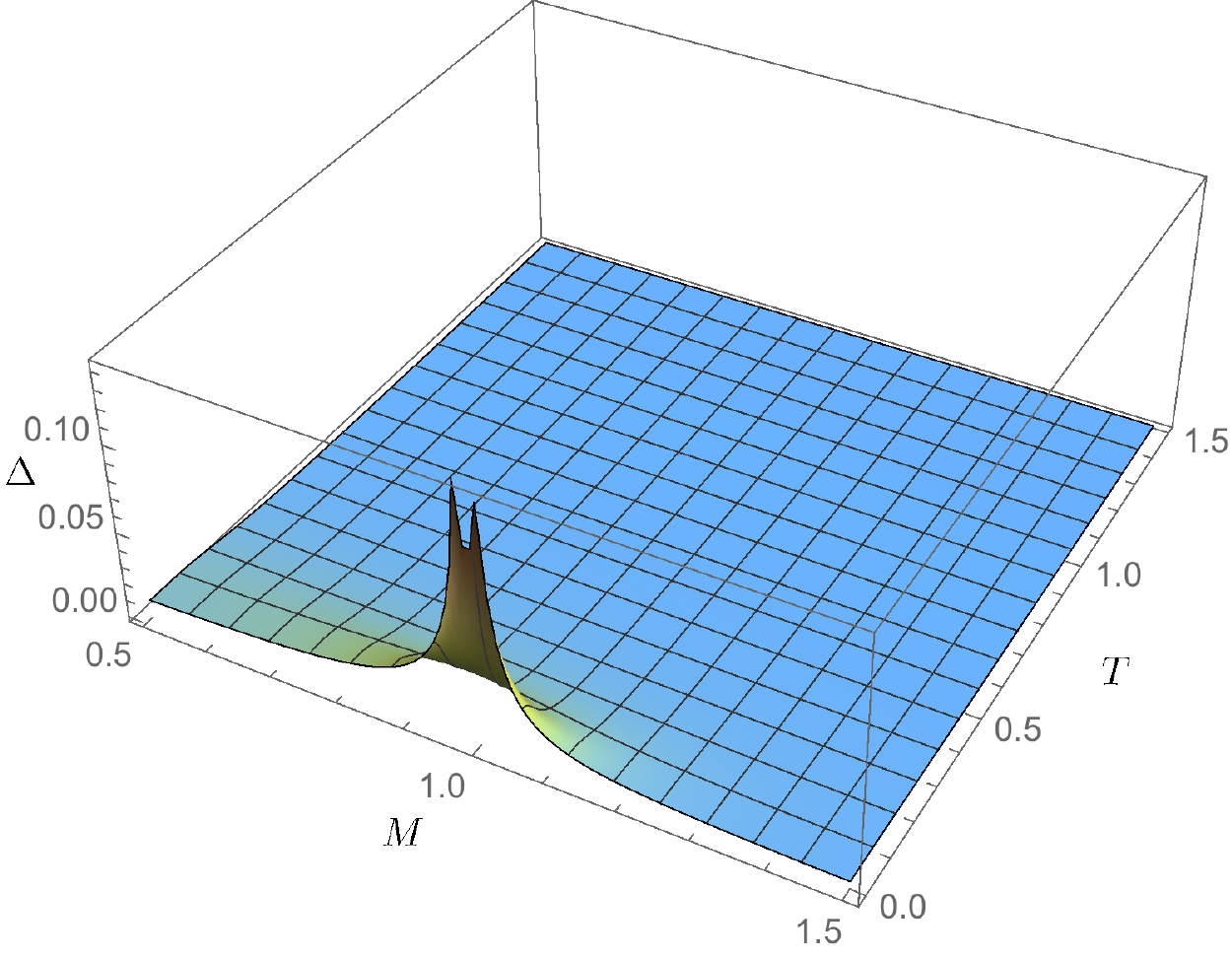}
\end{minipage}%
\caption{The fidelity for thermal states $\rho$, when probing the parameter of the Hamiltonian that drives the topological PT $\delta M =M'-M=0.01$ (left), and the temperature $\delta T=T'-T=0.01$ (center), and the Uhlmann connection, when probing the parameter of the Hamiltonian $M$ (right), for the Creutz ladder model (representative of the symmetry class AIII). The plot for $\Delta$ when deforming the thermal state along $T$ is omitted since it is equal to zero everywhere.}
\label{fig:fid}
\end{figure}
\end{widetext}

We see that for $T=0$ both fidelities exhibit a sudden drop in the neighbourhood of the gap-closing point $M=1$, signalling the topological quantum PT. As temperature increases, the drops of both fidelities at the quantum critical point are rapidly smoothened towards the $F=1$ value. This shows the absence~of both finite-temperature parameter-driven, as well as temperature-driven (i.e., thermal) PTs. The plot for $\Delta$, for the case $\delta T=0$, shows a behavior similar to that of the fidelity, while for $\delta M = 0$ we obtain no information, as $\Delta$ is identically equal to zero, due to the triviality of the Uhlmann connection associated to the mutually commuting states (a consequence of the Hamiltonian's independence on the temperature). $\Delta$ is sensitive to PTs for which the state change is accompanied by a change of the eigenbasis (in contrast to fidelity, which is sensitive to both changes of eigenvalues and eigenvectors). For TIs and TSCs, this corresponds to parameter-driven transitions only.

\begin{widetext}
 
\begin{figure}[h!]
\begin{minipage}{0.24\textwidth}
\includegraphics[width=0.8\textwidth,height=0.6\textwidth]{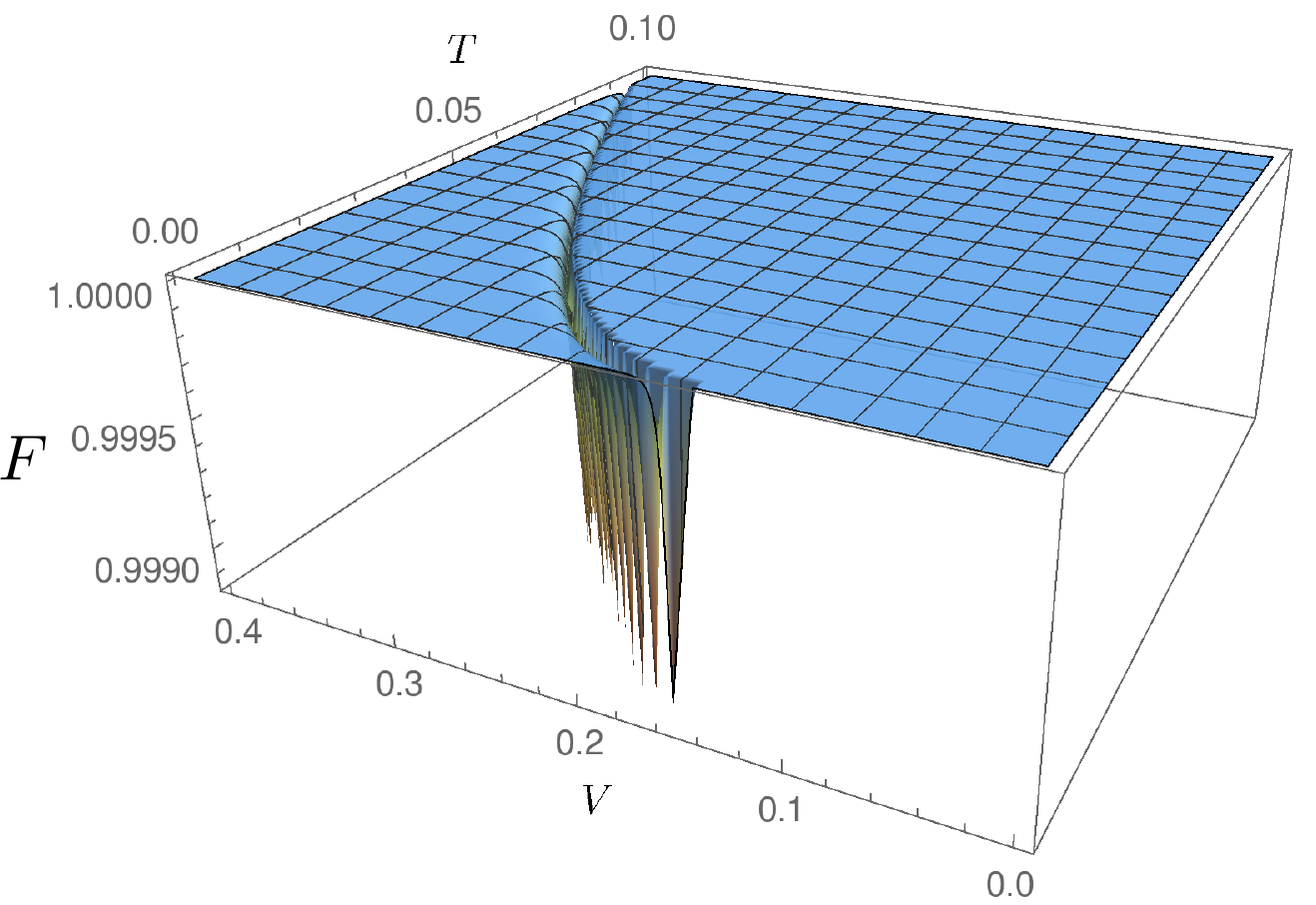}
\end{minipage}%
\begin{minipage}{0.24\textwidth}
\includegraphics[width=0.8\textwidth,height=0.6\textwidth]{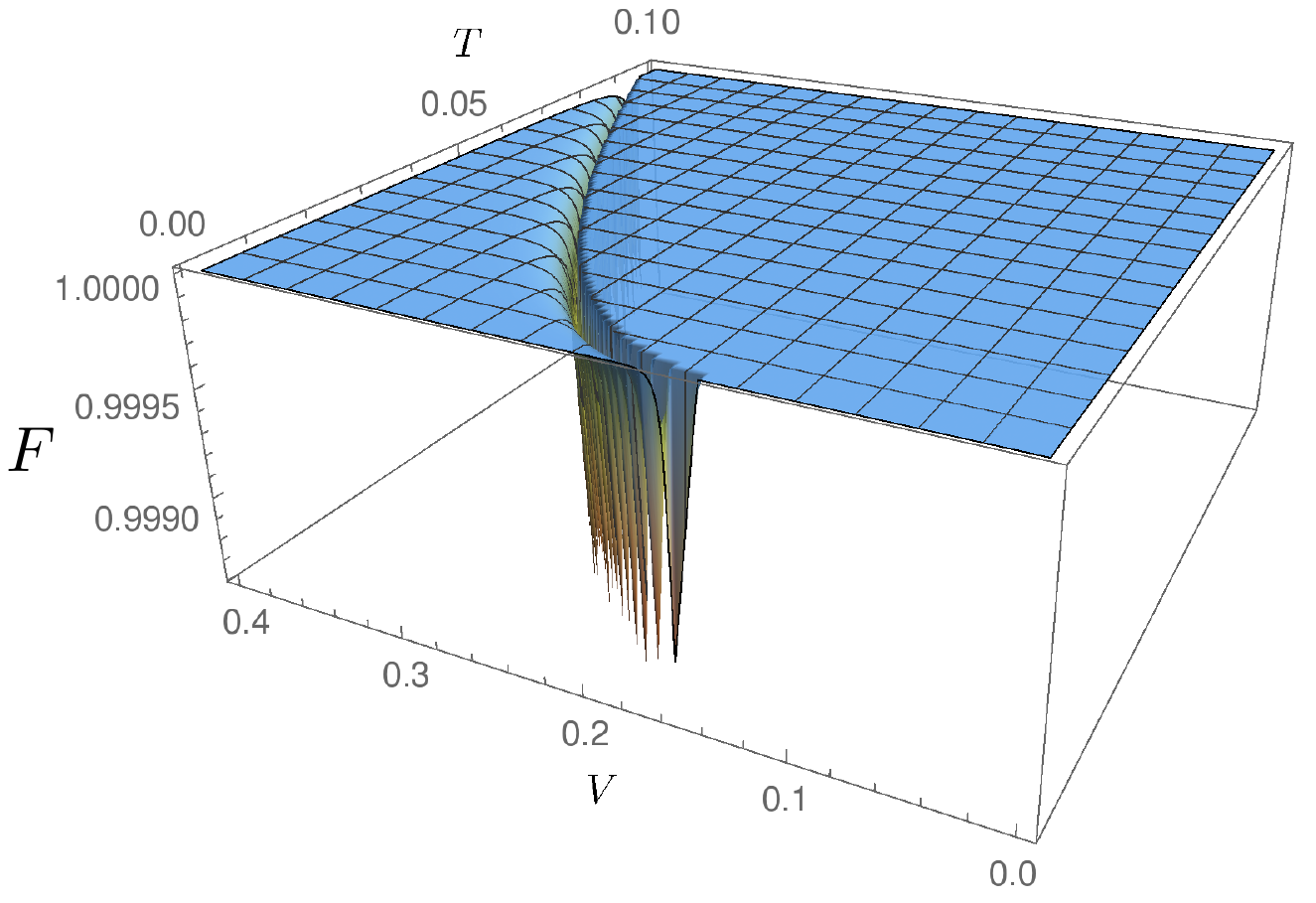}
\end{minipage}
\begin{minipage}{0.24\textwidth}
\includegraphics[width=0.8\textwidth,height=0.6\textwidth]{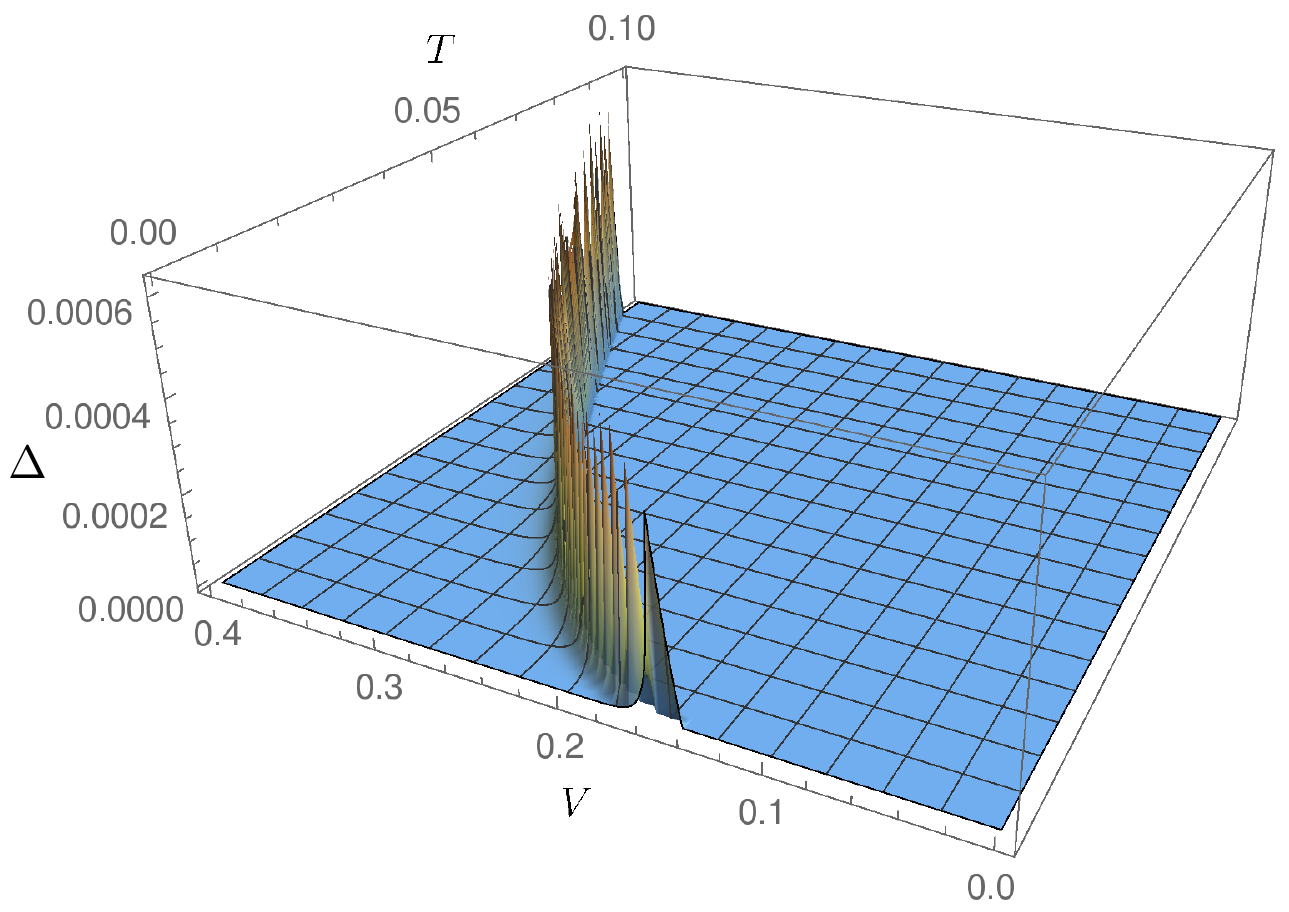}
\end{minipage}%
\begin{minipage}{0.24\textwidth}
\includegraphics[width=0.8\textwidth,height=0.6\textwidth]{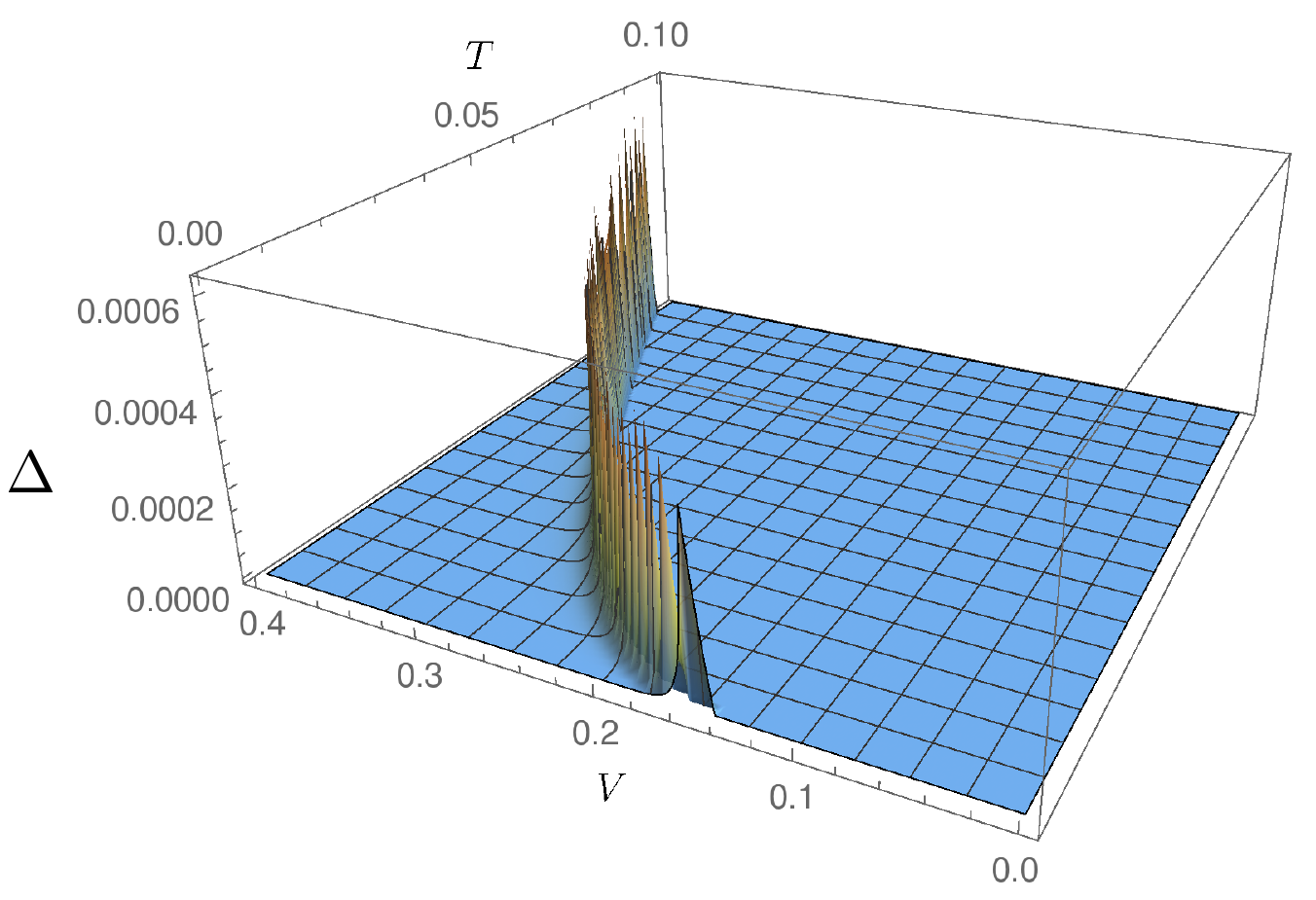}
\end{minipage}
\caption{The fidelity for thermal states $\rho$ when probing the parameter of the Hamiltonian $\delta V =V'-V=10^{-3}$ (left) and the temperature $\delta T=T'-T=10^{-3}$ (center left), and the Uhlmann connection (center right and right, respectively), for BCS superconductivity.}
\label{fig:bcs}
\end{figure}
\end{widetext}

We further study a topologically trivial superconducting system, given by the BCS theory, with the effective Hamiltonian 
\begin{align}
\mathcal{H}=\sum_{k} (\varepsilon_{k}-\mu)c_{k}^{\dagger}c_{k}-\Delta_{k} c_{k}^{\dagger}c_{-k}^{\dagger} + \text{H.c.},
\end{align}
where $\varepsilon_k$ is the energy spectrum, $\mu$ is the chemical potential, $\Delta_{k}$ is the superconducting gap, $c_{k}\equiv c_{k\uparrow}$ and $c_{-k}\equiv c_{-k \downarrow}$ are operators annihilating, respectively, an electron with momentum $k$ and spin up and an electron with momentum $-k$ and spin down. The gap parameter is determined in the above mean-field Hamiltonian through a self-consistent mass gap equation and it depends on the original Hamiltonian's coupling associated to the lattice-mediated pairing interaction $V,$ absorbed in $\Delta_k$ (for more details, see~\cite{pau:vie:08}). The solution of the equation renders the gap temperature-dependent.
In Fig.~\ref{fig:bcs}, we show the quantitative results for the fidelity and $\Delta$. We observe that both quantities show the existence of thermally driven PTs, as their abrupt change in the point of criticality at $T=0$, survive and drift, as temperature increases. Unlike TSCs, in this model the temperature does not only appear in the thermal state, but it is also a parameter of the effective Hamiltonian, resulting in the change of the system's eigenbasis and consequently nontrivial Uhlmann connection. For a detailed analysis and the explanation of the differences between the two, see SM4.

Finally, we also studied the behavior of the edge states for the TIs and the Majorana modes for the Kitaev chain, on open chains of 500 and 300 sites, respectively. In the case of TIs, we showed that the edge states localized at the boundary between two distinct topological phases, present at zero temperature, are gradually smeared out with the temperature increase, confirming the absence of temperature-driven PTs (see SM3.1 for detailed quantitative results and technical analysis). Our results on the edge states, obtained for systems in thermal equilibrium, agree with those concerning open systems treated within the Lindbladian approach~\cite{viy:riv:del:12} (and consequently, due to considerable computational hardness, obtained for an open chain of 8 sites). Similarly, we showed that the Majorana modes exhibit an abrupt change at the zero-temperature point of the quantum PT, while the finite-temperature behavior is smooth, confirming the results obtained through the fidelity and the Uhlmann connection analysis (see SM3.2). In the case of zero-temperature, Majorana modes of the Kitaev model are known to be good candidates for encoding qubit states in stable quantum memories, see~\cite{ali:12,ipp:riz:gio:maz:16} and references therein. The presence of robust Majorana modes at low, but finite, temperatures is a significant property which could be used in designing stable quantum memories in realistic setups.

\section*{Conclusions and Outlook}
\label{Conclusions and Outlook}

We studied the relationship between the fidelity and the Uhlmann connection over the system's \textit{parameter space} (including parameters of the system's Hamiltonian and temperature) and found that their behaviors are consistent whenever the variations of the parameters produce variations in the eigenbasis of the density matrix. By means of this analysis, we showed the absence of temperature-driven PTs in 1D TIs and TSCs. We clarified that the Uhlmann geometric phase considered in \textit{momentum space} is not adequate to infer such PTs, since it is only a part of the information contained in the Uhlmann holonomy. Indeed, this holonomy, as a function of temperature, is smooth (Eq. \eqref{eq:hol}), hence no PT-like phenomenon is expected. Furthermore, we performed the same analysis in the case of BCS superconductivity, where, in contrast to the former systems, thermally driven PTs occur and are captured by both the fidelity and the Uhlmann connection. This shows that, when changing the temperature, the density operator is changing both at the level of its spectrum and its eigenvectors. We analyzed in detail the origin of the differences between the BCS and the Kitaev chain and suggested that in realistic scenarios the gap of topological superconductors could also, generically, be temperature-dependent.

  Finally, we would like to point out possible future lines of research. The study of Majorana modes at finite temperature suggests that they can be used in achieving realistic quantum memories. This provides a relevant path for future research. Another related subject is to perform the same analysis using an open system approach where the system interacts with a bath and eventually thermalizes. There the parameter space would also include the parameters associated to the system-bath interaction.

\acknowledgements{B. M. and C. V. acknowledge the support from DP-PMI and FCT (Portugal) through the grants SFRH/BD/52244/2013 and PD/BD/52652/2014, respectively. B. M. acknowledges the support of Physics of Information and Quantum Technologies Group.  C. V. and N. P. acknowledge the support of SQIG -- Security and Quantum Information Group and UID/EEA/50008/2013. N. P. acknowledges the IT project QbigD funded by FCT PEst-OE/EEI/LA0008/2013. Support from FCT (Portugal) through Grant UID/CTM/04540/2013 is also acknowledged.
}

\section*{Supplemental Material}
\subsection{1. Analytic derivation of the closed expression for the fidelity}
\hspace{0.5cm} The fidelity between two states $\rho$ and $\rho'$ is given by
\begin{align}
F(\rho,\rho')=\text{Tr}\sqrt{\sqrt{\rho}\rho'\sqrt{\rho}}.
\end{align}
We consider unnormalized thermal states $\rho=\exp(-\beta H)$ and $\rho'=\exp(-\beta' H')$. At the end of the calculation one must, of course, normalize the expression appropriately. We wish to find closed expressions for the thermal states considered in the main text. In order to do that we will proceed by finding $e^C$, such that
\begin{align}
e^{A}e^{B}e^{A}=e^{C},
\end{align}
for $A=-\beta H$, $B=-\beta'H'$ and, ultimately, take the square root of the result. The previous equation is equivalent to
\begin{align}
e^{A}e^{B}=e^{C}e^{-A}.
\label{eq:1}
\end{align}
The Hamiltonians $H$ and $H'$ are taken to be of the form $\q{h}$, and thus we can write
\begin{align}
e^{A}=a_0+\q{a}, \nonumber \\
e^{B}=b_0+\q{b}, \nonumber \\
e^{C}=c_0+\q{c},
\end{align}
where all the coefficients are real, with the following constraints:
\begin{eqnarray}
\label{eq:c1}
\begin{cases}
& 1=\det e^{A}=a_{0}^2-\vec{a}^2,\\
\label{eq:c2}
& 1=\det e^{B}=b_{0}^2-\vec{b}^2,\\
\label{eq:c3}
& 1=\det e^{C}=c_{0}^2-\vec{c}^2,
\end{cases}
\end{eqnarray}
which are equivalent to $\tr A = \tr B = \tr C = 0$, since Pauli matrices are traceless. Let us proceed by expanding the LHS and the RHS of Eq.\eqref{eq:1},
\begin{align}
& (a_0+\q{a})(b_0+\q{b})=(c_0+\q{c})(a_0-\q{a}) \nonumber \\
& \Leftrightarrow a_0 b_0+ a_0\q{b}+\q{a}b_0+ (\q{a})(\q{b}) = c_0 a_0-c_0\q{a}+\q{c}a_0 -(\q{c})(\q{a})  \nonumber \\
& \Leftrightarrow a_0 b_0 +a_0 \q{b}+\q{a}b_0+ \vec{a}\cdot\vec{b}+i(\vec{a}\times \vec{b})\cdot \vec{\sigma}=c_0 a_0-c_0\q{a}+\q{c}a_0 -\vec{c}\cdot\vec{a}-i(\vec{c}\times \vec{a})\cdot\vec{\sigma}.
\end{align}
Now, collecting terms in $1$, $\vec{\sigma}$ and $i\vec{\sigma}$, we get a system of linear equations on $c_0$ and $\vec{c}$,
\begin{align}
\label{Eq:LS1}
\begin{cases}
& a_0 b_0+\vec{a}\cdot\vec{b}- a_0 c_0 +\vec{a}\cdot\vec{c}=0,\\
& a_0\vec{b}+b_0\vec{a}+\vec{a}c_0-a_0\vec{c}=0,\\
& \vec{a}\times \vec{b}-\vec{a}\times \vec{c}=0.
\end{cases}
\end{align}
The third equation from \eqref{Eq:LS1} can be written as $\vec{a}\times(\vec{b}-\vec{c})=0$, whose solution is given by $\vec{c}=\vec{b}+\lambda\vec{a}$, where $\lambda$ is a real number. This means that the solution depends only on two real parameters: $c_0$ and $\lambda$. Hence, we are left with a simpler system given by,
\begin{align}
\begin{cases}
& a_0 b_0+\vec{a}\cdot\vec{b}- a_0 c_0 +\vec{a}\cdot(\vec{b}+\lambda\vec{a})=0\\
& a_0\vec{b}+b_0\vec{a}+\vec{a}c_0-a_0(\vec{b}+\lambda \vec{a})=0
\end{cases}.
\end{align}
Or,
\begin{align}
\begin{cases}
& a_0 c_0 -\lambda\vec{a}^2=a_0 b_0+2\vec{a}\cdot\vec{b}\\
& (a_0\lambda-c_0)\vec{a}=b_0\vec{a}
\end{cases}.
\end{align}
In matrix form, the above system of equations can be written as
\begin{align}
\left[\begin{array}{cc}
a_0 & -\vec{a}^2\\
-1 & a_0	
\end{array}\right]\left[\begin{array}{c}
c_0\\
\lambda	
\end{array}
\right]=\left[\begin{array}{c}
a_0 b_0+2\vec{a}\cdot\vec{b}\\
b_0
\end{array}
\right].
\end{align}
Inverting the matrix, we get
\begin{align}
\left[\begin{array}{c}
c_0\\
\lambda	
\end{array}
\right]&=\frac{1}{a_0^2-\vec{a}^2}\left[\begin{array}{cc}
a_0 & \vec{a}^2\\
1 & a_0	
\end{array}\right]\left[\begin{array}{c}
a_0 b_0+2\vec{a}\cdot\vec{b}\\
b_0
\end{array}
\right] \nonumber \\
&=\left[\begin{array}{c}
(2 a_0 ^2-1) b_0+2 a_0\vec{a}\cdot\vec{b}\\
2(a_0 b_0+\vec{a}\cdot\vec{b})
\end{array}
\right],
\end{align}
where we used the constraints \eqref{eq:c1}. Because of the constraints, $c_0$ and $\lambda$ are not independent, namely,
$e^C=c_0+(\vec{b}+\lambda \vec{a})\cdot\vec{\sigma}$, and we get
\begin{align}
c_0^2-(\vec{b}+\lambda \vec{a})^2=c_0^2-\vec{b}^2-2\lambda \vec{a}\cdot \vec{b}-\vec{a}^2=1.
\end{align}
Now we want to make $A=-\beta H/2\equiv-\xi \vec{x}\cdot \vec\sigma/2$ and $B=-\beta 'H'\equiv-\zeta \vec{y}\cdot \vec\sigma$, with $\vec{x}^2=\vec{y}^2=1$ and $\xi$ and $\zeta$ real parameters, meaning,
\begin{align}
a_0=\cosh(\xi/2) \text{ and } \vec{a}=-\sinh(\xi/2) \vec{x},\\
b_0=\cosh(\zeta) \text{ and } \vec{b}=-\sinh(\zeta) \vec{y}.	
\end{align}
If we write $C=\rho \vec{z}\cdot \vec \sigma$ (because the product of matrices with determinant $1$ has to have determinant $1$, it has to be of this form),
\begin{align}
c_0 &=\cosh(\rho) \nonumber \\
&=(2 a_0 ^2-1) b_0+2 a_0\vec{a}\cdot\vec{b} \nonumber \\
&=(2\cosh ^2(\xi/2)-1)\cosh(\zeta)+2\cosh(\xi/2)\sinh(\xi/2)\sinh(\zeta)\vec{x}\cdot\vec{y} \nonumber \\
&=\cosh(\xi)\cosh(\zeta)+\sinh(\xi)\sinh(\zeta)\vec{x}\cdot\vec{y}.
\end{align}
For all the expressions concerning fidelity, we wish to compute $\text{Tr}(e^{C/2})=2\cosh(\rho/2)$. If we use the formula $\cosh(\rho/2)=\sqrt{(1+\cosh(\rho))/2}$, we obtain,
\begin{align}
\tr(e^{C/2})=2\sqrt{\frac{(1+\cosh(\xi)\cosh(\zeta)+\sinh(\xi)\sinh(\zeta)\vec{x}\cdot\vec{y})}{2}}.
\end{align}
Hence, if we let $\xi= \beta E/2$, $\vec{x}=\vec{n}$, $\zeta=\beta' E'/2$ and $\vec{y}=\vec{n}'$, then
\begin{align}
\tr(\sqrt{e^{-\beta H/2}e^{-\beta' H'}e^{-\beta H/2}})=2\sqrt{\frac{(1+\cosh(\beta E/2 )\cosh(\beta'E'/2)+\sinh(\beta E/2)\sinh(\beta'E'/2)\vec{n}\cdot\vec{n}')}{2}}.
\end{align}
To be able to compute the fidelities, we will just need the following expression relating the traces of quadratic many-body fermion Hamiltonians (preserving the number operator) and the single-particle sector Hamiltonian obtained by projection:
\begin{align}
\tr(e^{-\beta \mathcal{H}})=\tr(e^{-\beta \Psi^{\dagger}H\Psi})=\det(I+e^{-\beta H}).
\end{align}
From the previous results, it is straightforward to derive the following formulae for the fidelities concerning the BG states considered:\\

\begin{align}
	F(\rho,\rho')&=\prod_{k\in\mathcal{B}}\frac{\tr(e^{-\mathcal{C}_k/2})}{\tr(e^{-\beta \mathcal{H}_k})\tr(e^{-\beta' \mathcal{H}'_k})}\nonumber\\
	&=\prod_{k\in\mathcal{B}}\frac{\det(I+e^{-C_k/2})}{\det^{1/2}(I+e^{-\beta H_k})\det^{1/2}(I+e^{-\beta' H'_k})}\nonumber\\
	&=\prod_{k\in\mathcal{B}}\frac{2+\sqrt{2\left(1+\cosh(E_k/2T)\cosh(E'_k/2T')+\sinh(E_k/2T)\sinh(E'_k/2T')\vec{n}_k\cdot\vec{n}'_k\right)}}{\sqrt{(2+ 2\cosh (E_k/2T))(2+2\cosh (E'_k/2T'))}},\\
	\nonumber\\
	\end{align}
where the matrix $C_k$ is such that $e^{-C_k}=e^{-\beta H_k/2}e^{-\beta' H'_k}e^{-\beta H_k/2}$ and $\mathcal{C}_k=\Psi_k^{\dagger}C_k\Psi_k$ is the corresponding many-body quadratic operator. To compute $\Delta(\rho,\rho')$ one needs, in addition, $\tr \sqrt{\rho}\sqrt{\rho'}$. This can be done along the lines of what was done above, hence we shall omit the proof and just state the result: 
\begin{align}
\tr \sqrt{\rho}\sqrt{\rho'}=\prod_{k\in \mathcal{B}}\frac{2+2\left(\cosh(E_k/4T)\cosh(E'_k/4T')+\sinh(E_k/4T)\sinh(E'_k/4T')\vec{n}_k\cdot\vec{n}'_k\right)}{\sqrt{(2+ 2\cosh (E_k/2T))(2+2\cosh (E'_k/2T'))}}
\end{align}

\subsection{2. Results for the other models considered}
\subsubsection{i. Su-Schrieffer-Heeger (SSH)}
The Hamiltonian for the SSH model~\cite{su:sch:hee:79} is given by
\begin{align}
\mathcal{H}&=\sum_{i\in\mathbb{Z}}v c^{\dagger}_{i,A}c_{i,B}+w c^{\dagger}_{i,B}c_{i+1,A}+\text{H.c.},
\end{align}
where $c_i$ are fermionic annihilation operators, $A,B$ correspond to the two parts of the dimerized chain and $v,w$ are coupling constants. The change of the difference $|v-w|$ between the two parameters of the Hamiltonian drives the topological phase transition. In particular, the phase transition occurs for $|v-w|=0$.

\begin{figure}[h!]
\begin{minipage}{0.33\textwidth}
\includegraphics[width=0.7\textwidth,height=0.5\textwidth]{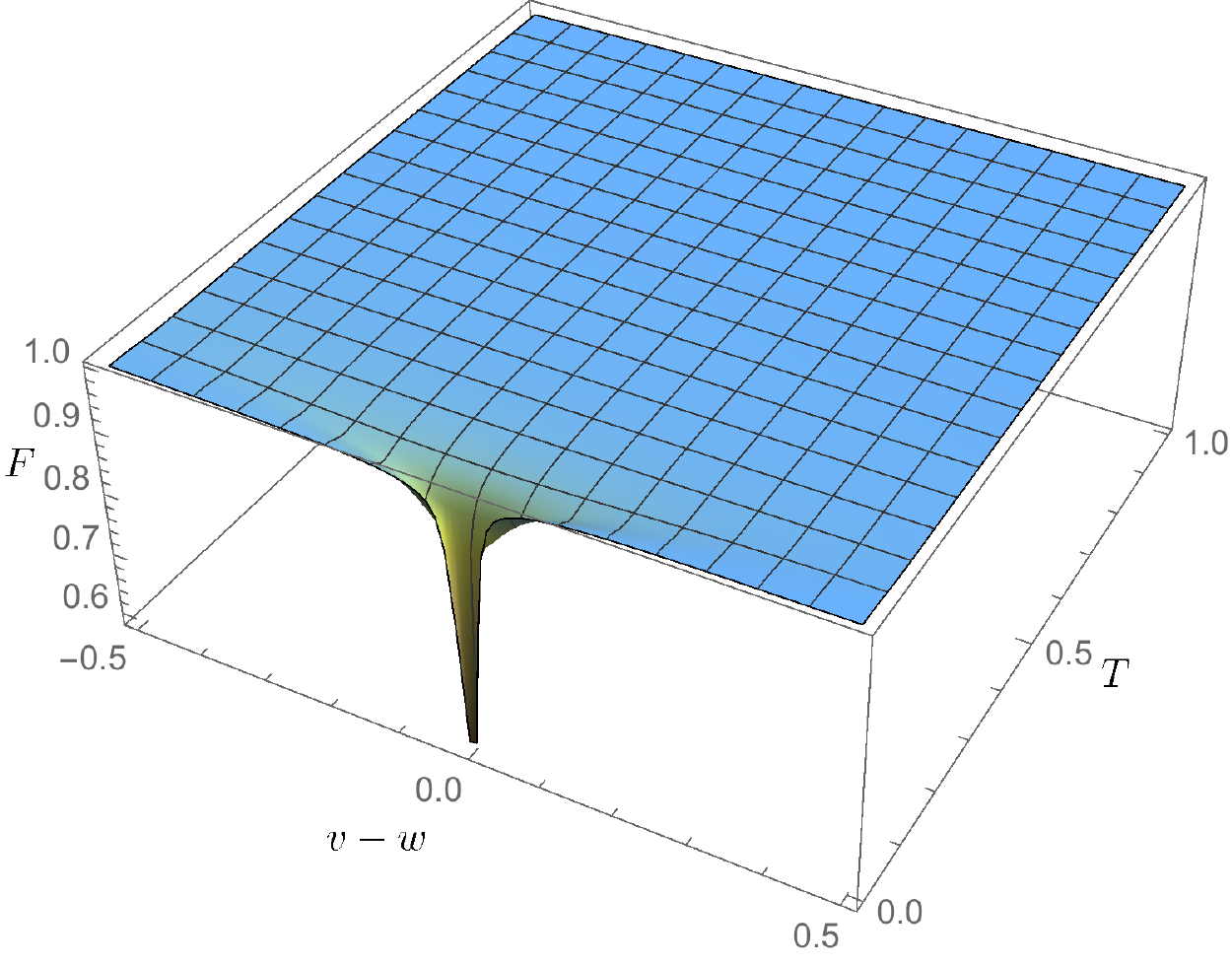}
\end{minipage}%
\begin{minipage}{0.33\textwidth}
\includegraphics[width=0.7\textwidth,height=0.5\textwidth]{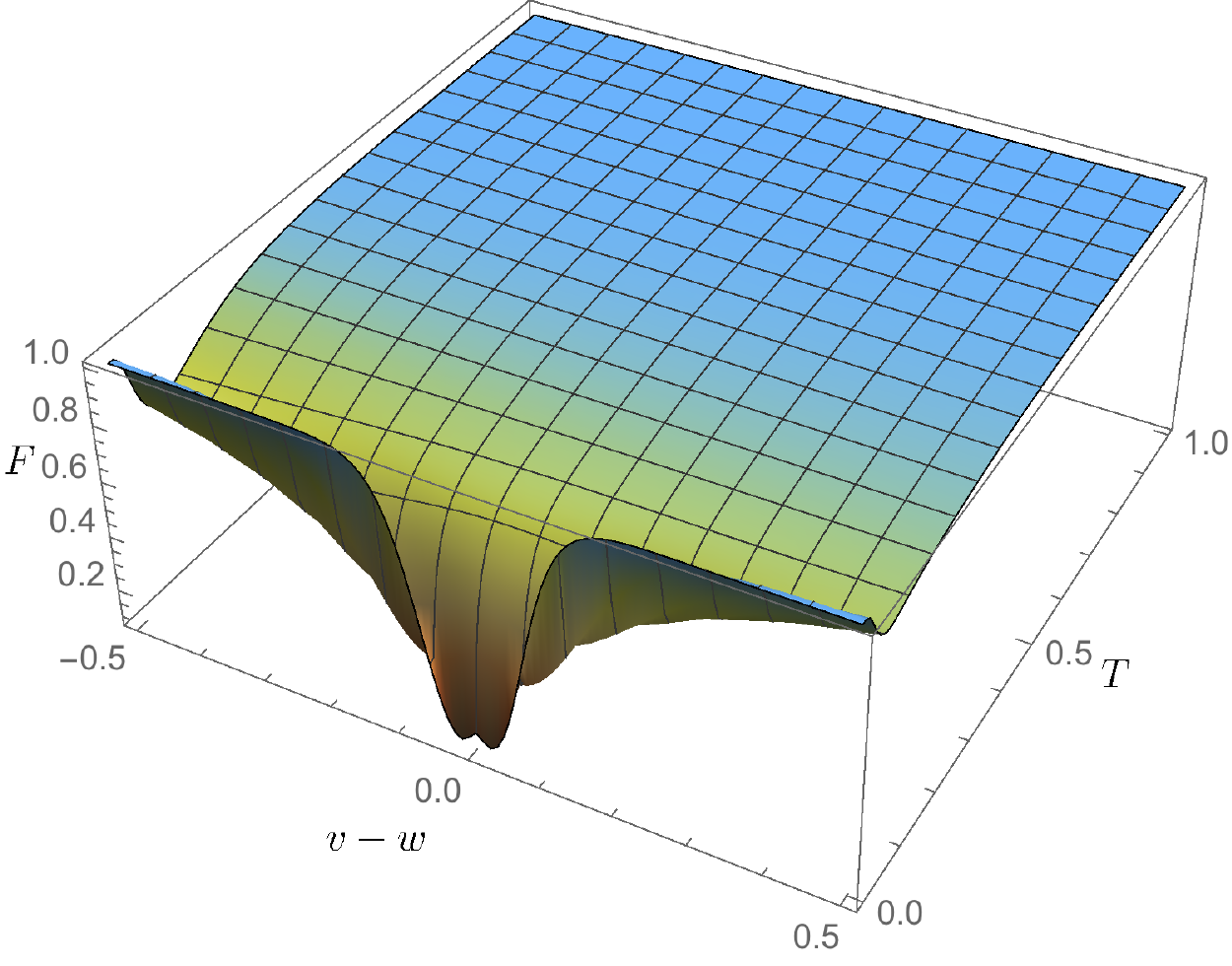}
\end{minipage}
\begin{minipage}{0.33\textwidth}
\includegraphics[width=0.7\textwidth,height=0.5\textwidth]{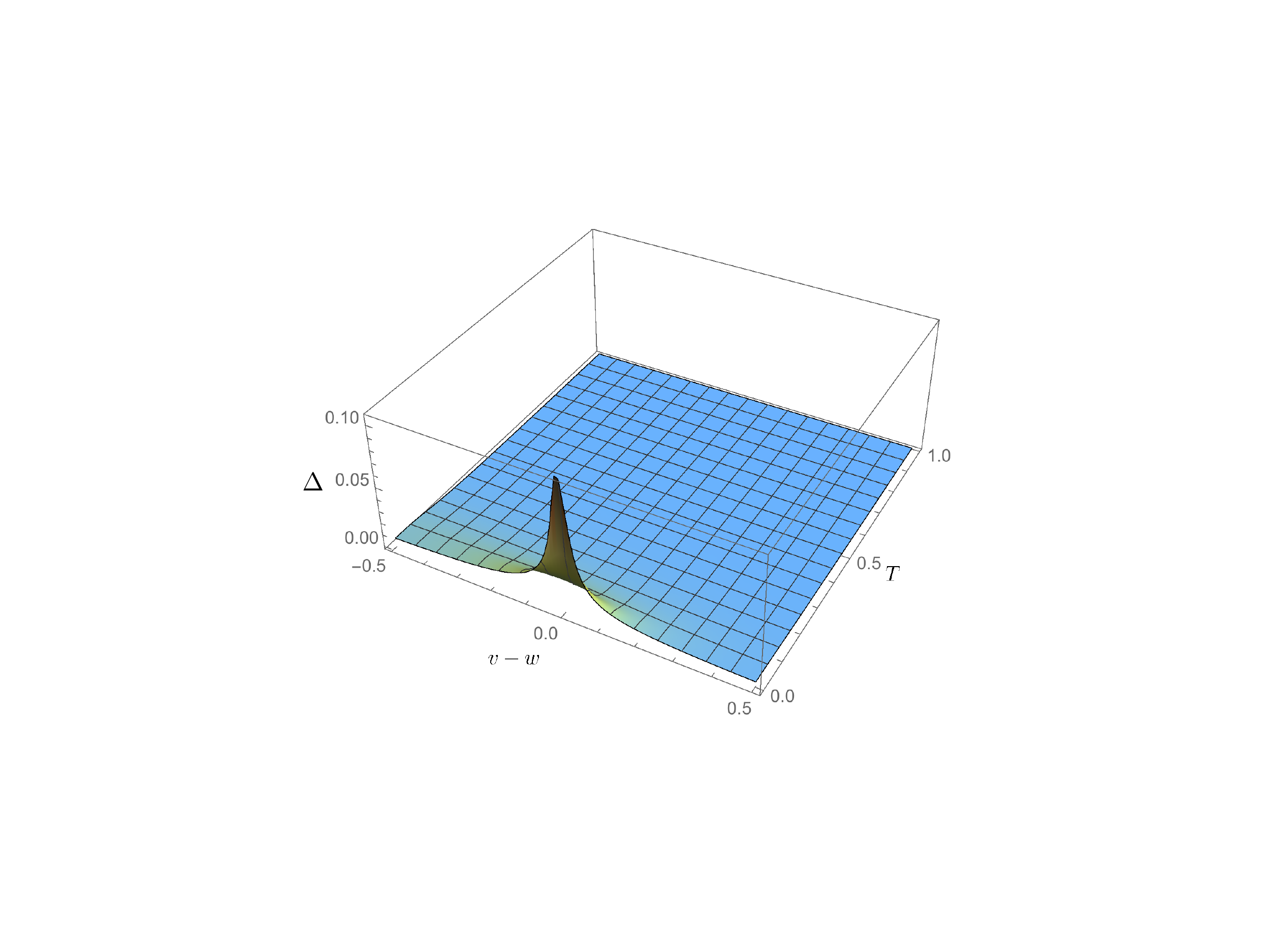}
\end{minipage}%
\caption{The fidelity for thermal states $\rho$, when probing the parameter of the Hamiltonian that drives the topological phase transition $\delta |v-w| =|v-w|'-|v-w|=0.01$ (left), and the temperature $\delta T=T'-T=0.01$ (center), and the Uhlmann connection, when probing the parameter of the Hamiltonian $|v-w|$ (right), for the SSH model (representative of the symmetry class BDI).}
\label{fig:fid}
\end{figure}

\subsubsection{ii. Kitaev Chain}
The Hamiltonian for the Kitaev Chain model~\cite{kit:cha:01} is given by
\begin{align}
\mathcal{H}&=-\mu\sum_{i=1}^N c^{\dagger}_i c_i+\sum_{i=1}^{N-1}\left[-t(c^{\dagger}_{i+1}c_i+c^{\dagger}_ic_{i+1})-|\Delta|(c_ic_{i+1}+c^{\dagger}_{i+1}c^{\dagger}_i)\right],
\end{align}
 where $\mu$ is the chemical potential, $t$ is the hopping amplitude and $\Delta$ is the superconducting gap. 
 We fix $t=0.5,\Delta=1$, while the change of $\mu$ along the sites of the line drives the topological phase transition. In particular, the phase transition occurs at $\mu=1$ (gap closing point).
 
 \begin{figure}[h!]
\begin{minipage}{0.33\textwidth}
\includegraphics[width=0.7\textwidth,height=0.5\textwidth]{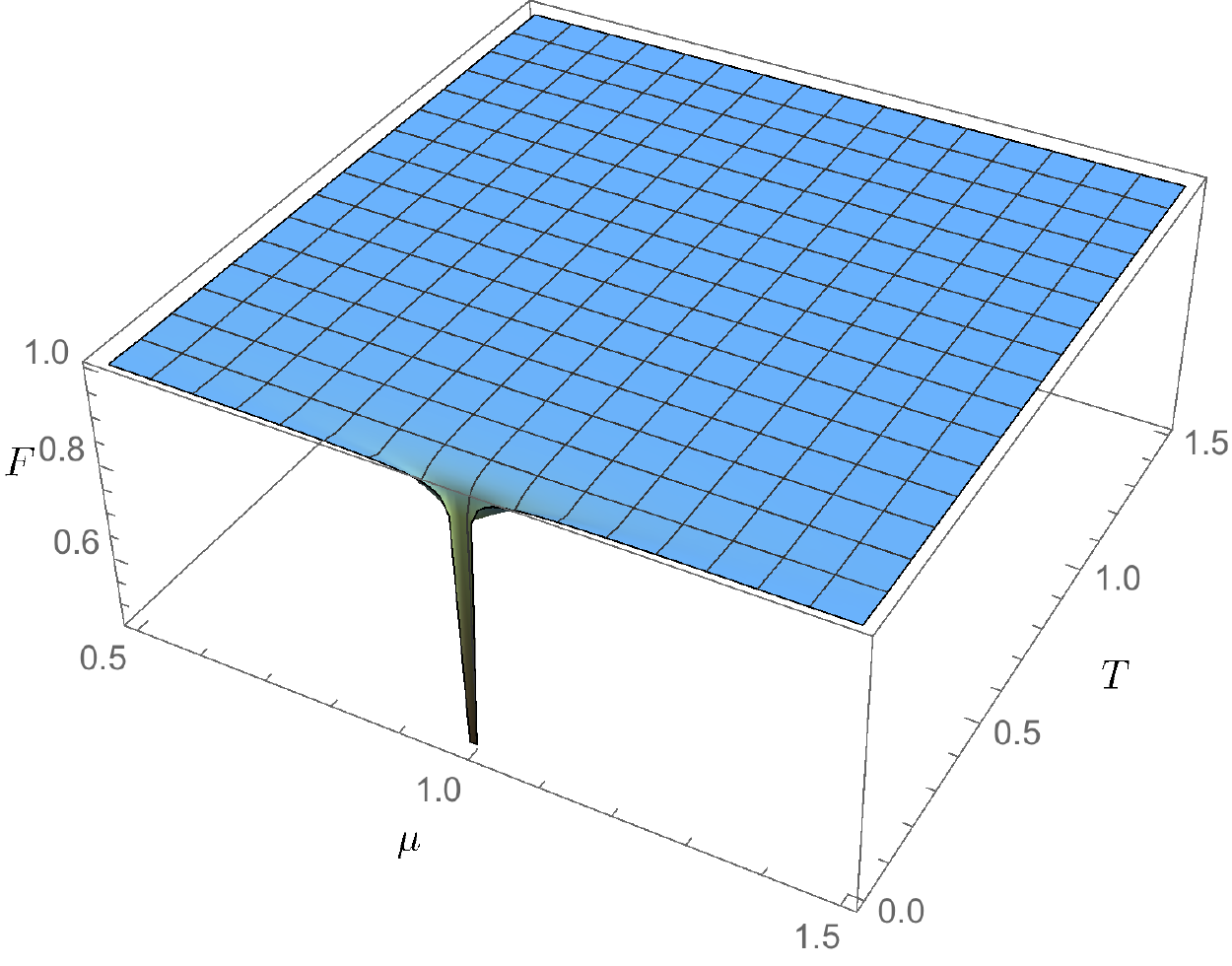}
\end{minipage}%
\begin{minipage}{0.33\textwidth}
\includegraphics[width=0.7\textwidth,height=0.5\textwidth]{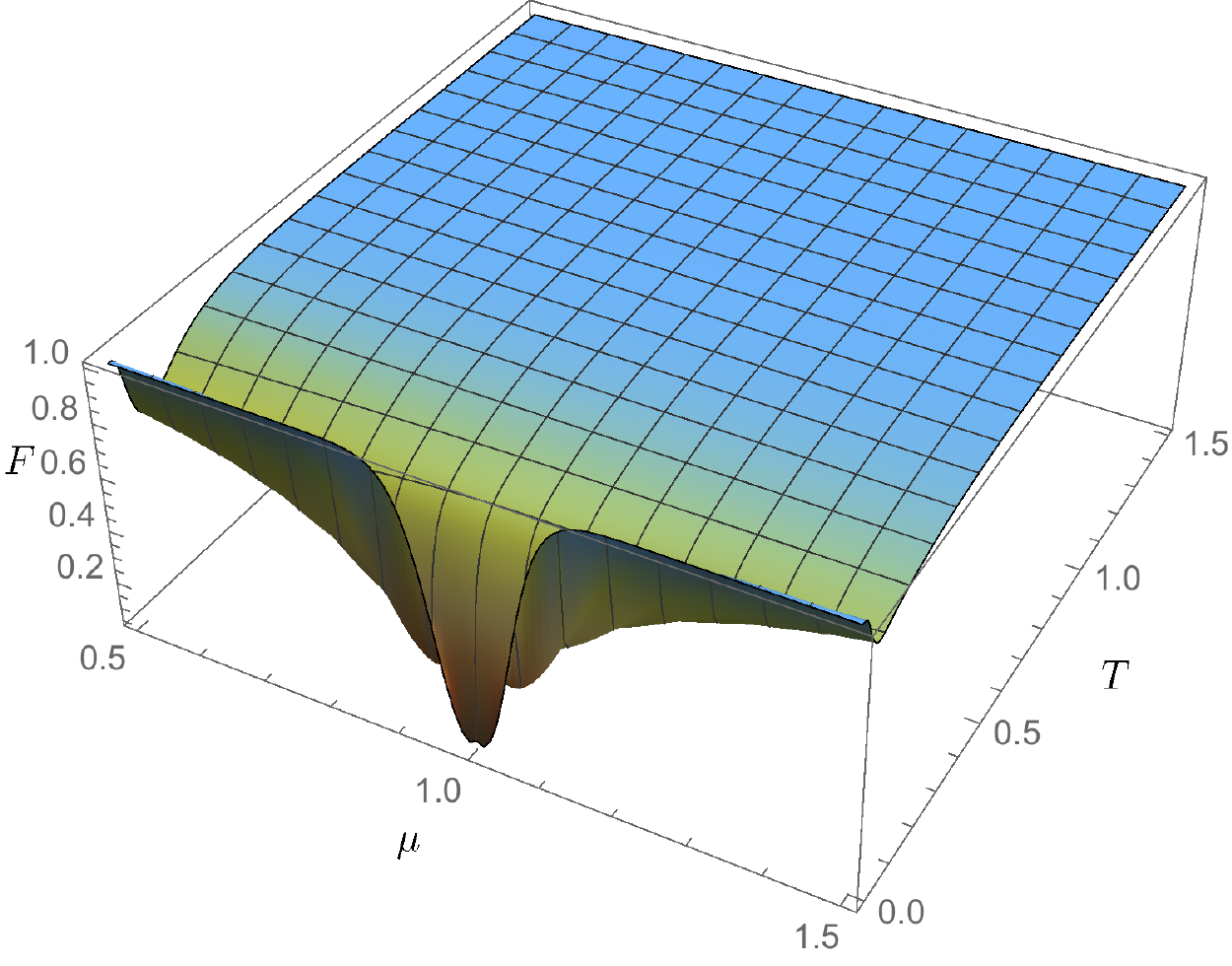}
\end{minipage}
\begin{minipage}{0.33\textwidth}
\includegraphics[width=0.7\textwidth,height=0.5\textwidth]{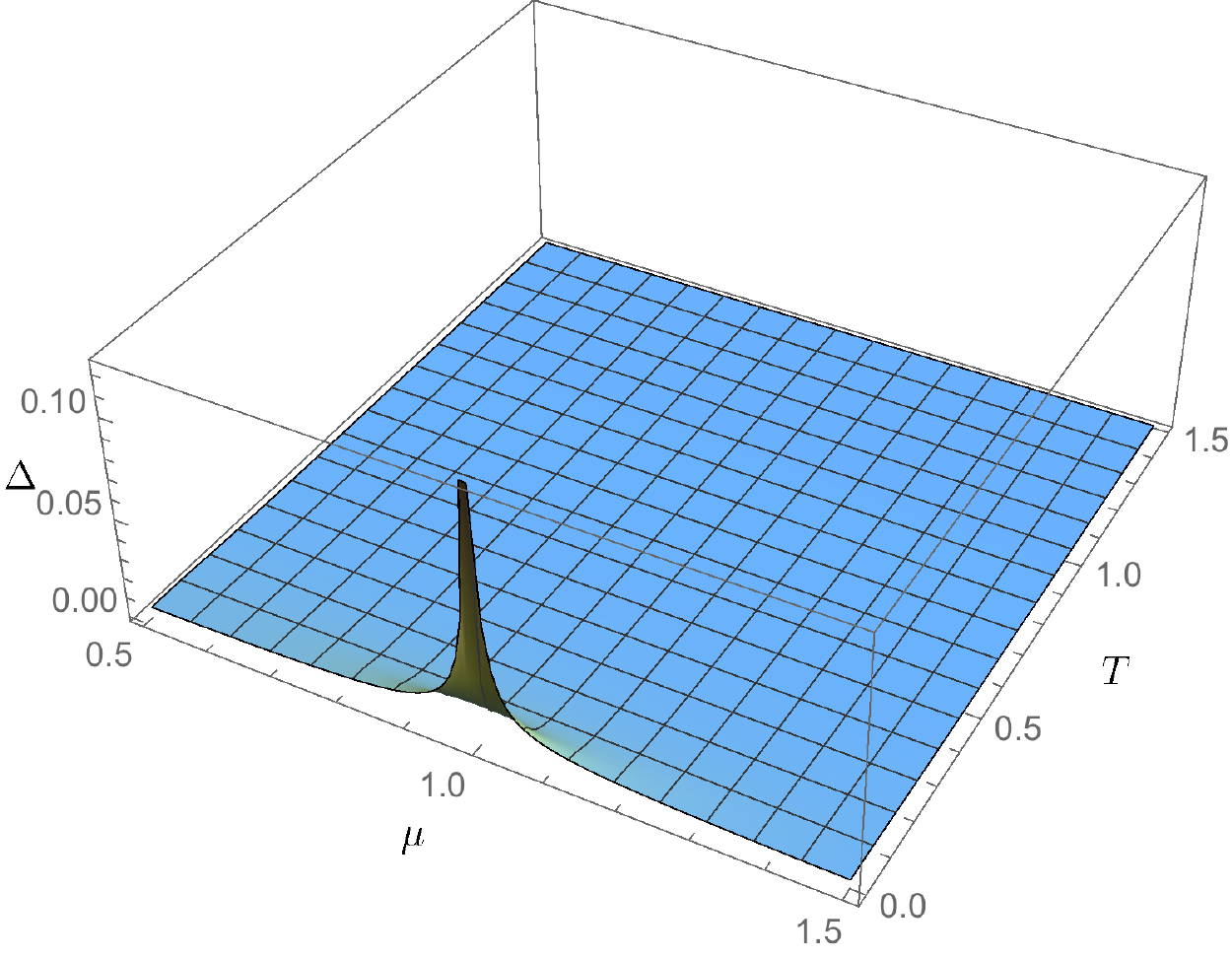}
\end{minipage}%
\caption{The fidelity for thermal states $\rho$, when probing the parameter of the Hamiltonian that drives the topological phase transition $\delta \mu =\mu'-\mu=0.01$ (left), and the temperature $\delta T=T'-T=0.01$ (center), and the Uhlmann connection, when probing the parameter of the Hamiltonian $\mu$ (right), for the Kitaev chain model (topological superconductor).}
\label{fig:fid}
\end{figure}

\subsection{3. The edge states}

For the case of the Creutz ladder, the Su-Schrieffer-Heeger (SSH) model and the Kitaev chain, when considering the system on a finite-size chain with open boundary conditions, the bulk-to-boundary principle predicts the existence of zero modes localized at the ends of the chain, whenever the bulk is in a topologically non-trivial phase. It is then possible to consider the associated thermal states, $\rho=\exp(-\beta \mathcal{H})/Z$, and probe the effects of temperature. The study of the Uhlmann connection and the fidelity conducted for the above models suggests that at zero temperature the edge states should exhibit an abrupt change as the system passes the point of quantum phase transition, while at finite temperatures they should smoothly change, slowly being washed away with the temperature increase, as a consequence of the absence of finite-temperature transitions. Below, we first study topological insulators in Subsection~3.1, while in Subsection~3.2 we analyze a  topological superconductor given by the Kitaev model, showing the agreement with the above inferred behavior.

\subsubsection{3.1 Topological Insulators}
\begin{figure}[h!]
\begin{minipage}{0.45\textwidth}
\centering
\includegraphics[scale=0.5]{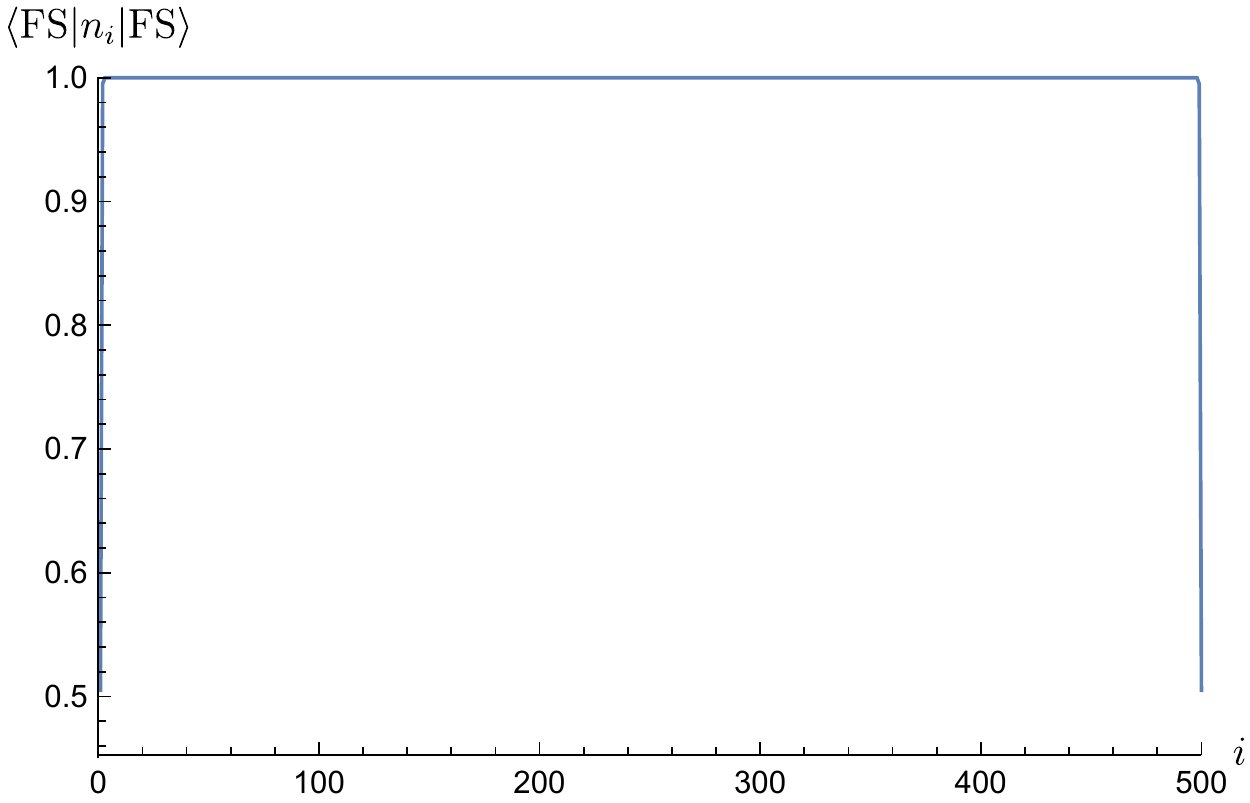}
\end{minipage}
\begin{minipage}{0.45\textwidth}
\centering
\includegraphics[scale=0.5]{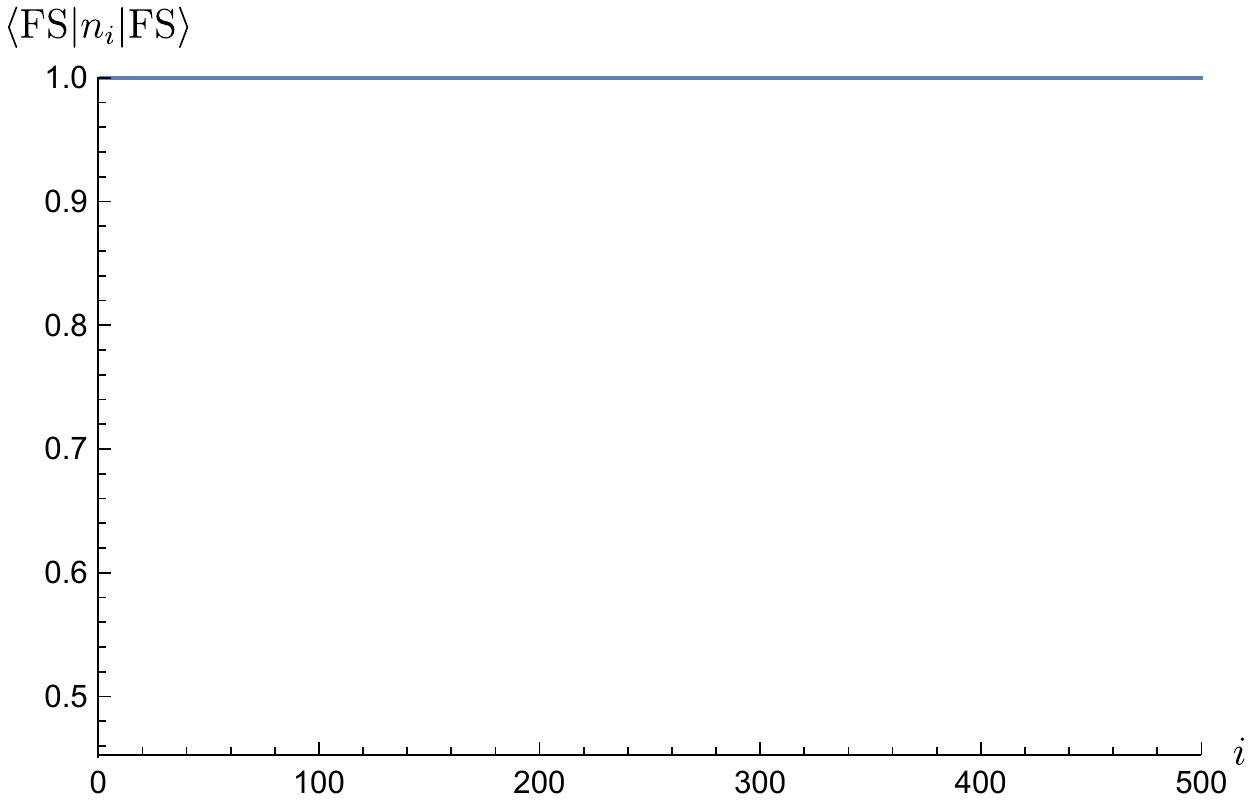}
\end{minipage}
\caption{Fermi sea expectation value of the occupation number $n_i=a_i^\dagger a_i+b_i^{\dagger}b_i$ as a function of position $i$ on a chain of 500 sites with open boundary conditions. On the left panel the system is in a topologically non-trivial phase with $2K=1,M=0.1,\phi=\pi/2$. On the right panel the system is in a topologically trivial phase with $2K=1,M=1.0001,\phi=\pi/2$.}
\label{fig:FS_occupation_number}
\end{figure}

 Let us consider the Creutz ladder model as a representative of topological insulators. In the trivial phase, the spectrum decomposes into two bands of states separated by a gap. At zero chemical potential, the zero-temperature limit of $\rho$ is the projector onto the Fermi sea state $\ket{\text{FS}},$ obtained by occupying the lower band. On a topologically non-trivial phase, however, the spectrum is composed of the two bands {\em and} the zero modes. At zero chemical potential, the zero temperature limit of $\rho$ is now the projector onto the ground state manifold of $\mathcal{H}$, which is spanned by $\ket{\text{FS}}$ and additional linearly independent states by creating excitations associated to the zero modes. Since the Fermi sea does not have these edge state excitations (exponentially localized at the boundary) included, the occupation number as a function of position, $n_i=a_i^\dagger a_i+b_i^{\dagger}b_i$, will see this effect at the boundary of the chain. Indeed, this is what we see in Fig.~\ref{fig:FS_occupation_number}: the occupation number, as a function of position, drops significantly at the edges in the topologically non-trivial phase. On the other hand, on the topologically trivial side the occupation number stays constant throughout the whole chain (both bulk and the edges).

If we want the thermal state's $T=0$ limit to be the Fermi sea, we have to add a very small (negative) chemical potential. It has to be small enough so that the lower band gets completely filled. In the following Fig.~\ref{fig:BG_occupation_number}, we see that the expectation value $\tr(\rho n_i)$ coincides with $\bra{\text{FS}}n_i\ket{\text{FS}}$ in the $T=0$ limit and that the deviation of the occupation number at the edge from that in the bulk gets washed out smoothly as the temperature increases. In fact, in the large temperature limit, the state is totally mixed, implying that the expected value of the occupation number will be constant and equal to $1$, as a function of position. Similar situation occurs in the SSH model. We see that the above results in Figs.~\ref{fig:FS_occupation_number} and~\ref{fig:BG_occupation_number} coincide with the results obtained by the fidelity analysis. Moreover, they, too, confirm the results based on the study of the Uhlmann connection in terms of the quantity $\Delta$. 

\begin{figure}[h!]
\begin{minipage}{0.45\textwidth}
\centering
\includegraphics[scale=0.55]{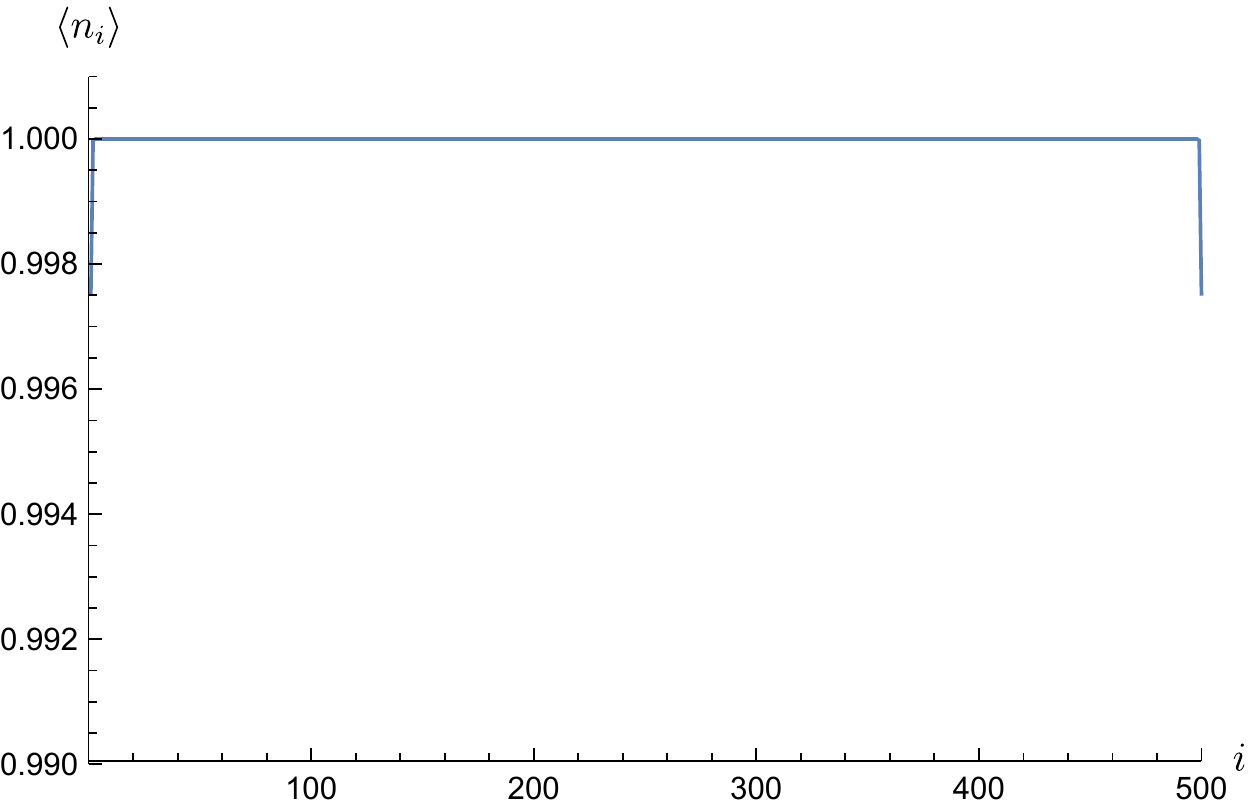}
\includegraphics[scale=0.55]{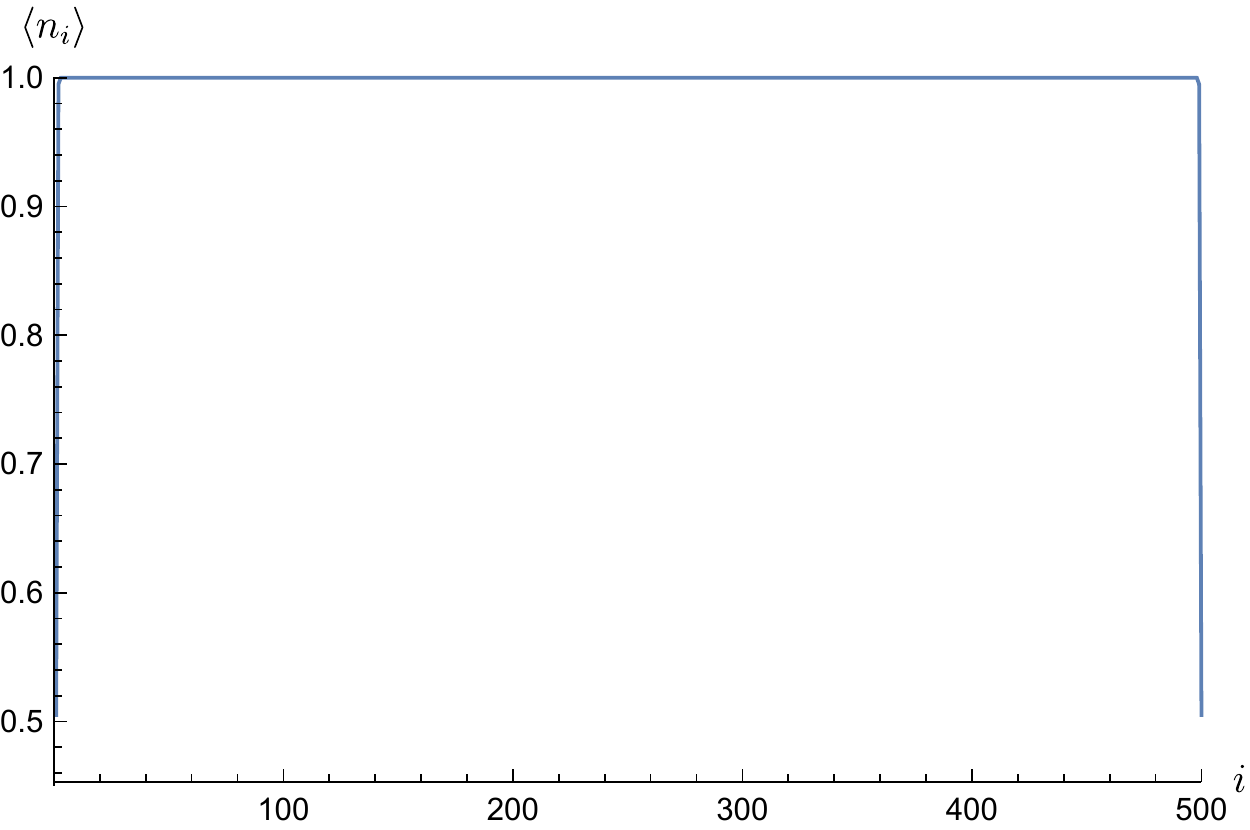}
\end{minipage}
\begin{minipage}{0.45\textwidth}
\centering
\includegraphics[scale=0.55]{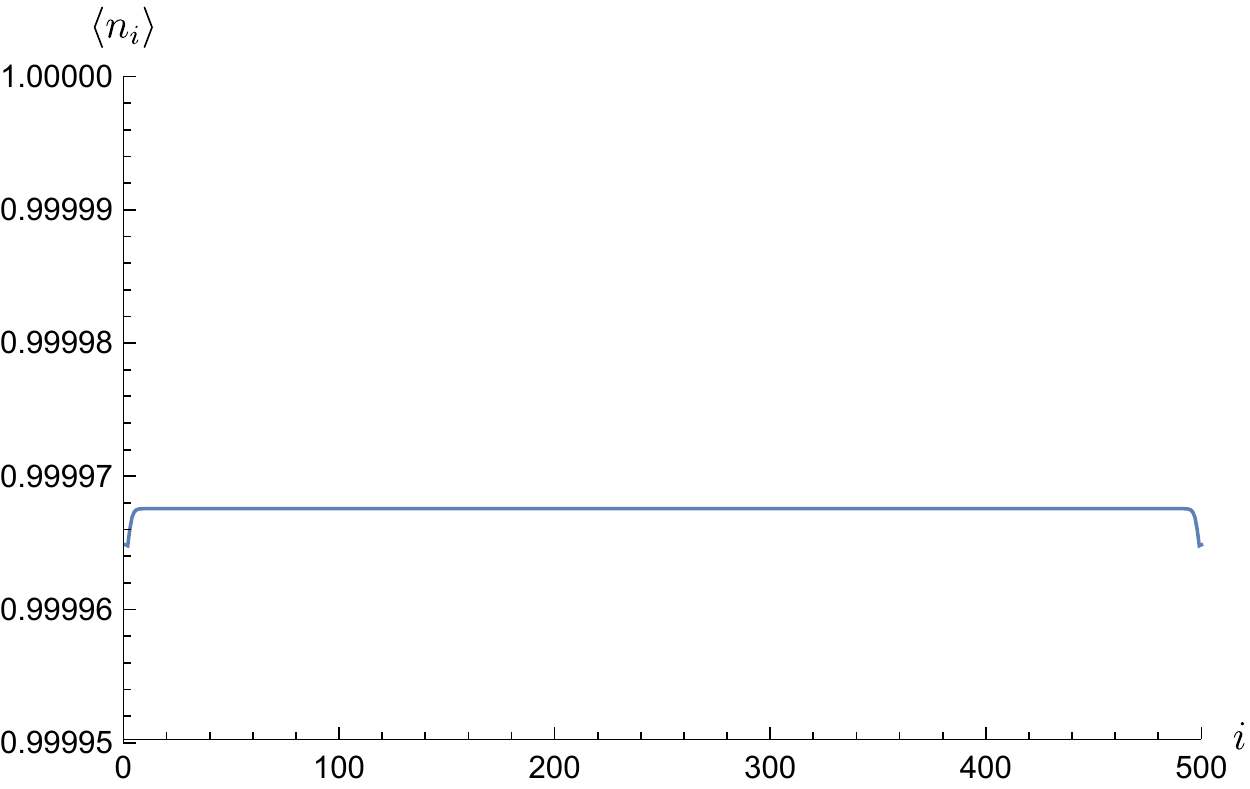}
\includegraphics[scale=0.55]{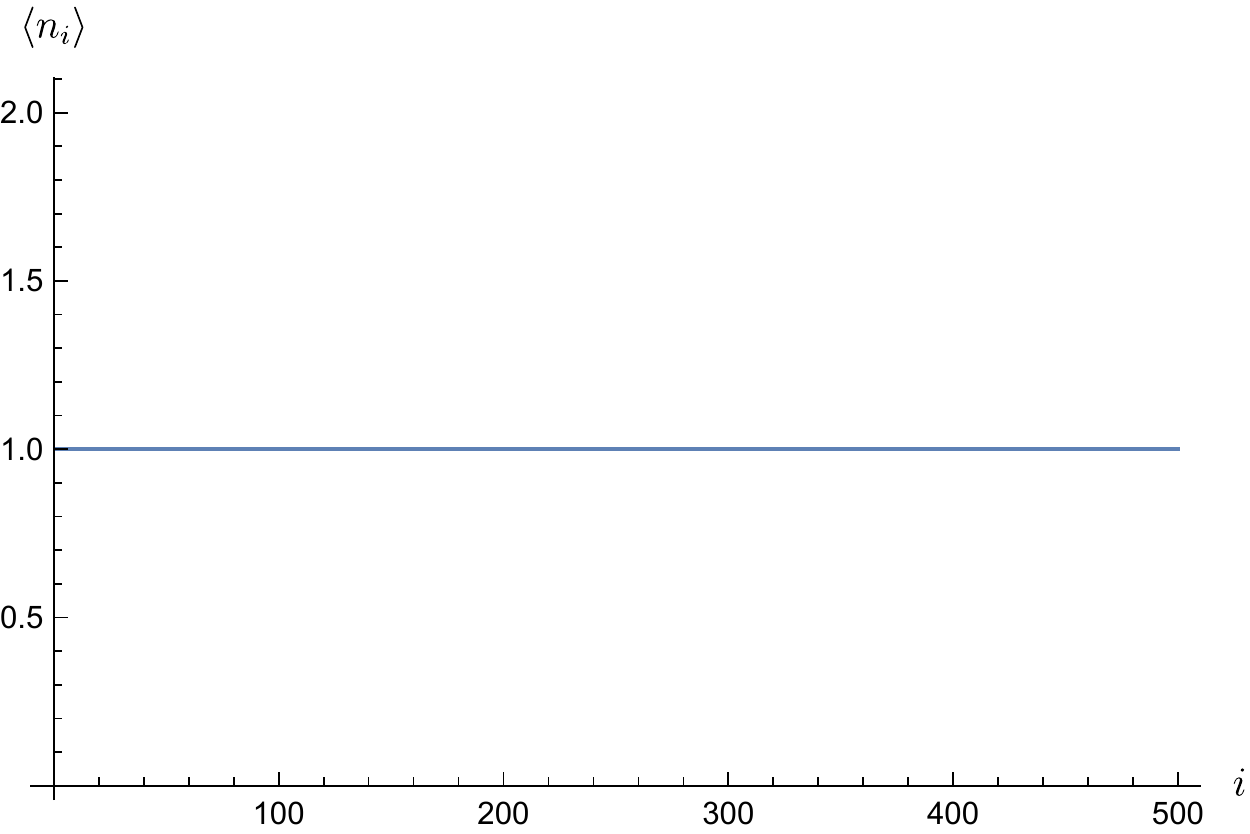}
\end{minipage}
\caption{Expectation value of the occupation number $n_i=a_i^\dagger a_i+b_i^{\dagger}b_i$ as a function of position $i$ on a chain of 500 sites with open boundary conditions, in a topologically non-trivial phase with $2K=1,M=0.1,\phi=\pi/2$ (left panel), for temperatures $T=10^{-5}$ (down) and $T=0.2$ (up). On the right panel we have a topologically trivial phase near the critical value of the parameter $2K=1,M=1.0001,\phi=\pi/2$, for temperatures $T=10^{-5}$ (down) and $T=0.2$ (up). Increasing $M$, the edge behavior is washed out smoothly, for finite $T$, and it becomes trivial as for the $T=0$ case.}
\label{fig:BG_occupation_number}
\end{figure}

\subsubsection{3.2 Topological Superconductors}

 As far as the superconducting Kitaev model is concerned, the chemical potential is a parameter of the Hamiltonian and we cannot lift the zero modes from the zero-temperature limit of $\rho$ with the above method. Moreover, the Kitaev Hamiltonian does not conserve the particle number, and adding chemical potential associated to the total particle number would not lift the zero modes even if $\mu$ were not a parameter of the Hamiltonian.  Note though, that the total number of Bogoliubov {\em quasi-particles} which diagonalize the Hamiltonian is conserved. Hence, we add a very small (negative) chemical potential associated with the total quasi-particle number, thereby lifting the Majorana zero modes energy. Notice that in the case of topological insulators this procedure coincides with the one applied above: since the Hamiltonian conserves the total particle number, and the quasi-particle creation operators are linear combinations of {\em just} the particle creation operators (and not of the holes as well), the total quasi-particle and particle numbers coincide. We found that the good quantity to be studied is not the occupation number as a function of the position in the chain, but the ratio between the average particle occupation number at the edge and the average particle occupation number at the bulk $f(\mu;T) = \langle n_{\text{edge}}\rangle/\langle n_{\text{bulk}}\rangle$ (without loss of generality, we have chosen for $n_{\text{bulk}}$ the site in the middle of the chain, since it is approximately constant throughout the bulk). 
 In Fig. 5, we present the results obtained for a chain with open boundary conditions, consisting of 300 sites.

 \begin{figure}[h!]

\centering
\includegraphics[scale=0.6]{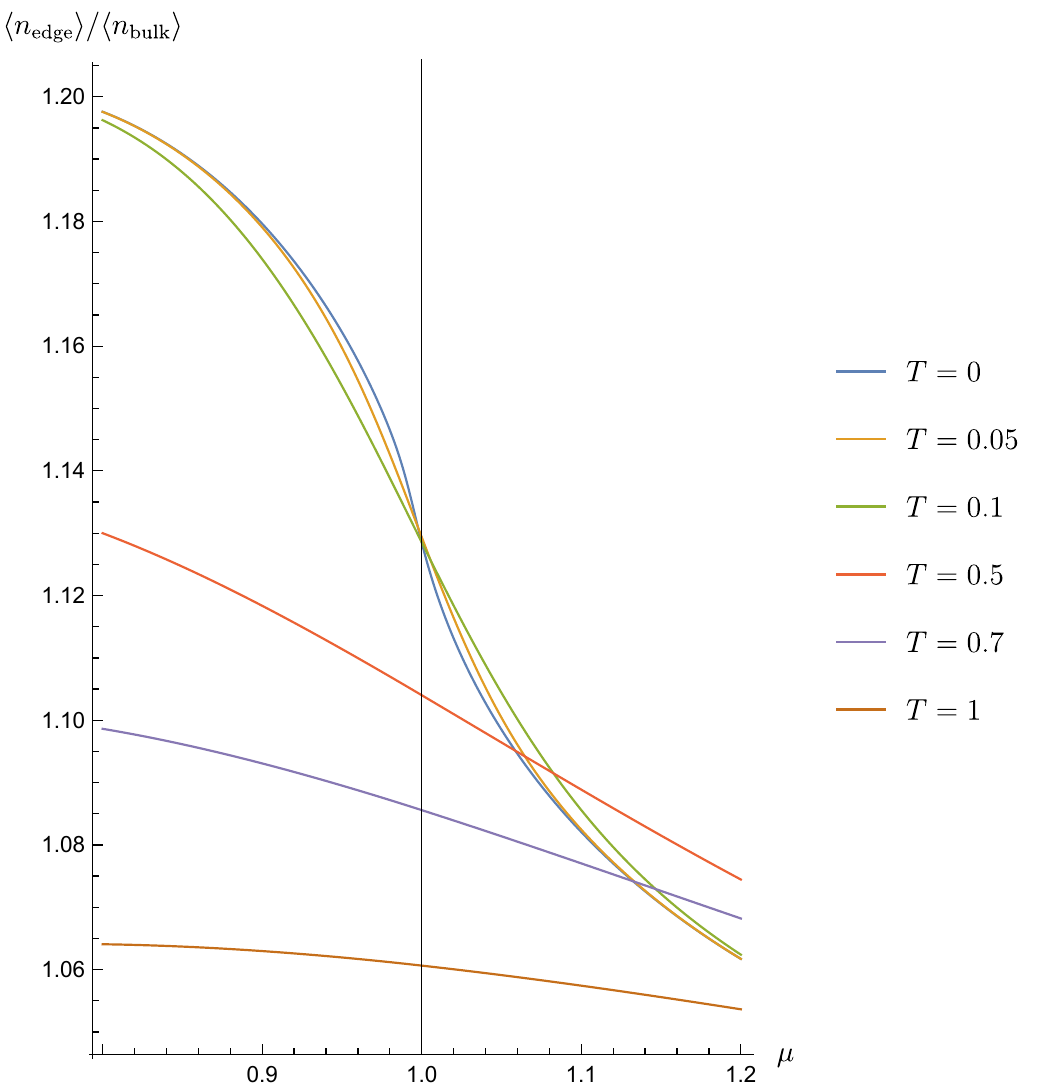}

\caption{$\langle n_{\text{edge}}\rangle/\langle n_{\text{bulk}}\rangle$ as a function of the chemical potential $\mu$ for a chain of 300 sites with open boundary conditions, for several values of the temperature $T$.}
\label{fig:majorana}
\end{figure}

The results are consistent with the behavior inferred by the Uhlmann connection (and the fidelity): Majorana modes exhibit an abrupt change at zero temperature (a signature of the quantum phase transition), while for fixed finite temperatures they smoothly change with the parameter change, and are slowly washed away with the temperature increase. Indeed, the behaviour of the finite-temperature curves is smooth, while the zero-temperature quench-like curve is expected to develop a discontinuity at $\mu = 1$ in the thermodynamic limit (see for example Fig.4(b) and the respective discussion in~\cite{qua:zur:10}). To show this more accurately, one needs considerably higher computational power to probe chain lengths of much higher orders of magnitude, a relevant future direction of work. 

We note that this new method to study Majorana modes is more general and also applicable in the case of insulators. The results obtained for topological insulators using this new method lead to the same qualitative conclusion regarding the behavior of the edge states (consistent with our previous results). In order to avoid repetition, we omit presenting the respective results.

In general, the behavior of the edge states and the associated Majorana modes show an interesting property of the systems which, at finite temperatures, despite the absence of phase transitions, they keep exhibiting their topological features even on the ``trivial side'' of the phase diagram (for parameter values for which on zero temperature the system is topologically trivial). At zero temperature, the Majorana modes are known to be good candidates for qubit encoding, thus they could be used to achieve fault-tolerant quantum computation, see~\cite{ali:12,ipp:riz:gio:maz:16} and references therein. Therefore, the aforementioned property of Majorana modes at the low but finite-temperature regime, is potentially significant in constructing stable quantum memories in more realistic scenarios. Furthermore, the existence of stable quantum memories has considerable impact in cryptography~\cite{rod:mat:pau:sou:17,pir:ott:spe:wee:bra:llo:geh:jac:and:15,ber:chr:col:ren:ren:10,lou:alm:and:pin:mat:pau:14,lou:ars:pau:pop:prv:16}.

 \subsection{4. The Uhlmann connection and the temperature-driven transitions: BCS vs. Kitaev chain}
 
The study of the Uhlmann connection at finite temperatures showed, both for topological insulators (Creutz ladder and SSH models), as well as for topological superconductors (Kitaev model), that there exist no finite-temperature transitions driven by the Hamiltonians' parameter(s). 
 
Regarding the finite-temperature transitions driven by the temperature, as stated in the manuscript, the Uhlmann connection quantifies the rate of change of a system's eigenvectors. Since this is one of the main points of our work, we here briefly recall the argument presented in the second section (third paragraph) of the main text. Indeed, if the two states $\rho$ and $\rho'$ commute, $[\rho , \rho']=0$, we have $F(\rho , \rho') = \mbox{Tr}|\sqrt{\rho}\sqrt{\rho'}| = \mbox{Tr}\sqrt{\rho}\sqrt{\rho'}$. Thus, the Uhlmann factor $U$ associated to the two states is trivial, $U=I$, and consequently $\Delta(\rho , \rho') = \mbox{Tr}|\sqrt{\rho}\sqrt{\rho'}| -\mbox{Tr}\sqrt{\rho}\sqrt{\rho'} = 0$. We see that the Uhlmann connection, and in particular the quantity $\Delta$, quantifying the rate of change of the system's eigenbasis, reflects the quantum contribution to the state distinguishability (see also~\cite{zan:ven:gio:07}, equation (3), in which the Bures metric is split into classical and non-classical terms, the first quantifying the change of system's eigenvalues and the second the change of the corresponding eigenvectors). Thus, the Uhlmann connection is trivial in the cases of the topological systems considered, as their Hamiltonians do not explicitly depend on temperature and thus commute with each other.

On the other hand, the mean-field BCS Hamiltonian considered in our paper does explicitly depend on the temperature. Indeed, as the results presented in the Results section of the main paper clearly show, the change of the eigenbasis of the BCS thermal states carries the signature of a thermally driven phase transition (see the bottom right plot of FIG.~2). Note that, having a purely non-classical contribution, such a temperature driven transition has quantum features as well, which is on its own an interesting consequence of the study of the Uhlmann connection.

It is an interesting question to compare the two cases of superconductors studied, and try to isolate the reasons for such difference. To see why the two cases are different, note that in the case of the BCS superconductivity we consider the mean-field Hamiltonian 
\begin{equation}
\label{eq:BCS_MF}
\mathcal{H}=\sum_{k} (\varepsilon_{k}-\mu)c_{k}^{\dagger}c_{k}-\Delta_{k} c_{k}^{\dagger}c_{-k}^{\dagger} + \text{H.c.},
\end{equation}
in which the gap $\Delta(V,T)$ is a function of temperature. Had we considered a more fundamental ``pairing Hamiltonian',' which takes into account the quartic electron interaction mediated by the phonons of the lattice,
\begin{equation}
\label{eq:BCS_P}
\mathcal{H}^{P}=\sum_{k} (\varepsilon_{k}-\mu)c_{k}^{\dagger}c_{k}-\sum_{k,k'} V_{k,k'}c_{k'}^{\dagger}c_{-k'}^{\dagger}c_{-k}c_{k} + \text{H.c.},	
\end{equation}
the Uhlmann connection would be trivial. The mean-field Hamiltonian is obtained by the BCS decoupling scheme with an averaging procedure setting the effective gap for the mean-field state $\rho = e^{-\beta\mathcal{H}}/Z$ (for simplicity, we assume $V_{kk'} = -V$ for $k,k'$ close to the Fermi momentum $k_F$, and zero otherwise) to be
\begin{equation}
\label{eq:gap}
\Delta(V,T) = -\sum_{k'} V_{kk'} \langle c_{-k}c_{k}\rangle = V \sum_k \mbox{Tr}(c_{-k}c_{k}\rho ).	
\end{equation}
In other words, the effective mean-field Hamiltonian~\eqref{eq:BCS_MF} is obtained from~\eqref{eq:BCS_P} by expanding $c_{-k}c_{k} = \langle c_{-k}c_{k}\rangle + \delta(c_{-k}c_{k})$ around the suitably chosen superconducting ground state. Thus, $\mathcal{H}$ breaks the $\mbox{U}(1)$ particle-number conservation symmetry of $\mathcal{H}^P$ to a residual $\mathbb Z_2$ symmetry, to accommodate the superconducting properties of the system. As a result, the Uhlmann connection becomes sensitive to temperature-driven transitions as well.

On the other hand, the Hamiltonian of the Kitaev model is phenomenological, modelled upon the success of the related BCS mean-field Hamiltonian. In this model, the gap is, for simplicity, considered to be temperature-independent. It would be interesting to probe this assumption in experiments with realistic topological superconducting materials. Our method based on the Uhlmann connection could then be useful in the analysis of such experiments.

The above discussion shows an interesting consequence of our analysis in terms of the Uhlmann connection. The temperature dependence of the BCS mean-field gap originally came as a consequence of a particular formal decoupling scheme. One might thus question whether a gap of a general superconducting material should also a priori depend on the temperature. Our analysis shows that a generic superconducting Hamiltonian of the form of~\eqref{eq:BCS_MF} would exhibit temperature-driven transitions, accompanied by the enhanced state distinguishability in terms of the system's eigenbasis, only in the case that the gap does depend on the temperature.

\bibliographystyle{unsrt}
\bibliography{Many_body.bib}

\end{document}